\documentclass[apjfonts]{emulateapj}

\slugcomment{ApJ, accepted 30th June, 2010}

\shorttitle{Particle transport in Protoplanetary Disks}
\shortauthors{Hughes \& Armitage}

\begin{document}

\title{Particle Transport in Evolving Protoplanetary Disks: \\ Implications for Results from {\em Stardust}}

\author{Anna L.~H. Hughes\altaffilmark{1,2} and Philip J. Armitage\altaffilmark{1,2}}
\altaffiltext{1}{JILA, 440 UCB, University of Colorado, Boulder, CO~80309-0440}
\altaffiltext{2}{Department of Astrophysical and Planetary Sciences, University of Colorado, Boulder}

\begin{abstract}
Samples returned from comet 81P/Wild~2 by the {\em Stardust} mission confirm that substantial quantities of crystalline silicates were incorporated into the comet at the time of its formation. We investigate the constraints that this observation places upon protoplanetary disk physics, under the assumption that outward transport of particles processed at high temperatures occurs via a combination of advection and turbulent diffusion in an evolving disk. We also look for possible constraints on the formation locations of such particles. Our results are based upon one-dimensional disk models that evolve with time under the action of viscosity and photoevaporative mass loss, and track solid transport using an ensemble of individual particle trajectories. We find that two broad classes of disk model are consistent with the {\em Stardust} findings. One class of models features a high particle diffusivity (a Schmidt number ${\rm Sc} < 1$), which suffices to diffuse particles up to 20~$\mu$m in size outward against the mean gas flow. For ${\rm Sc} \ge 1$ such models are unlikely to be viable, and significant outward transport appears to require that the particles of interest settle into a midplane layer that experiences an outward gas flow. In either class of models, the mass of inner disk material that reaches the outer disk is a strong function of the initial compactness of the disk. Hence, models of grain transport within steady-state disks underestimate the efficiency of outward transport. Neither model results in sustained outward transport of very large particles exceeding a millimeter in size.  We show that in most circumstances, the transport efficiency falls off rapidly with time. Hence, high-temperature material must be rapidly incorporated into icy bodies to avoid fallback to small radii. We suggest that significant radial transport may only occur during the initial phase of rapid disk evolution.  It may also vary substantially between disks depending upon their initial mass distributions.  We discuss how our model may inform recent {\em Spitzer} observations of crystalline silicates in T Tauri star-disk systems.
\end{abstract}


\section{Introduction}
\label{Introduction}

Comets are believed to be some of the most primitive bodies in the Solar System.  As such, they preserve information about Solar nebula conditions during the early stages of planet formation.  Samples collected from the comet 81P/Wild~2 during the course of the {\em Stardust} mission show that the nonvolatile material of the comet is rich in relatively large grains ($\sim$few--20 $\mu$m) of crystalline silicates \citep{Brownleeetal2006,Zolenskyetal2006,Westphaletal2009}. Crystalline silicates are not detected in the diffuse interstellar medium (ISM) and, although there is debate as to the formation mechanism of these grains in protoplanetary disks, they are not thought to form as far out in the disk as comet 81P/Wild~2 is believed to have originated. The {\em Stardust} results thus provide direct evidence for the outward transport of particles through protoplanetary disks. Spectroscopic observations suggest that similar processes may be widespread in T~Tauri disks. {\em Spitzer} studies have revealed an apparent link between disk crystallinity, and grain growth and settling, but otherwise these disks exhibit a wide range of crystalline-to-amorphous silicate mass ratios and spatial distributions. This is suggestive of a great diversity in protoplanetary dust processing \citep{vanBoekeletal2005,Watsonetal2009,Olofssonetal2009}.

Because no process has been proposed to form crystalline silicates out as far as the comet-forming regions, a mechanism is needed that can explain the presence of particles processed through high temperatures in cold regions of the Solar nebula. One obvious possibility is radial mixing induced by turbulence. Turbulence is probably necessary for gas within the disk to accrete \citep{ShakuraSunyaev1973}, and that same turbulence will result in diffusion of gaseous tracers and small particles coupled to the gas by aerodynamic forces \citep{MorfillVolk1984,ClarkePringle1988}. It is unclear, however, whether turbulent transport suffices to explain the observations, or whether other physical processes are also needed. Examples of such additional mechanisms  include the ballistic launching of particles in a wind from near the inner edge of the disk \citep{Shuetal2001}, photophoretic gas-pressure forces acting on grains in an optically thin disk \citep{KraussWurm2005,Mousisetal2007}, and radiation pressure on larger grains near the surface of the disk \citep{Vinkovic2009}. 

The radial transport of both gaseous and particle species within a  turbulent disk has been studied by several authors \citep{Gail2001,BockeleeMorvanetal2002,KellerGail2004,Boss2004,DullemondApaiWalch2006,Ciesla2007,AlexanderArmitage2007,Garaud2007,Boss2008,Dullemondetal2008,Ciesla2009,TurnerCarballidoSano2010}.  Here, we use a particle-based approach to model the advection and turbulent transport of non-interacting dust grains within evolving protoplanetary disks. Our goals are to identify the conditions under which significant outward transport of particles can occur, bearing in mind the range of uncertainty in disk physics and evolution. We quantify the extent of radial transport as a function of the size of the grains, the initial compactness of the disk, the relative diffusivity of disk-gas tracers, and on the vertical profile of the gas flow and particle distribution.  Initially compact configurations that expand rapidly appear to have the greatest promise to explain the {\em Stardust} results, though unambiguous predictions require accurate knowledge of the internal disk-flow structure and of crystalline-silicate formation mechanisms.  We find that the importance of the advection of solids within the gas flow means that the outward transport efficiency drops significantly for larger particles ($\sim$few millimeters), and at later times, thereby limiting the extent of mixing uniformity achievable within the disk.

In Section \ref{Background} we provide a summary of the relevant observations of silicates in both primitive Solar System bodies and in other protoplanetary disks.  In Section \ref{Disk}  we describe our modeling approach, detailing our 1D gas-disk evolution and particle transport models. In Section \ref{Overview}, we step through a set of examples designed to illustrate the basic effects of various model parameters.  In Section \ref{Results}, we present and analyze results for a baseline case of particle transport in two scenarios of radial gas flow, then consider the effects of varying the size of the particles, the diffusivity, and the initial compactness of the disk.  We present our conclusions in Section \ref{Conclusions}.  

\section{Observational constraints}
\label{Background}

Observations of dust in the diffuse ISM show that the vast majority of silicates there are amorphous, with sub-micron grain sizes and a generally balanced Mg-Fe composition \citep{MolsterKemper2005,Woodenetal2007}.  Therefore, the crystalline silicates that we observe in the Solar System and in disks around other stars are expected to have formed after the onset of the star-disk--system formation. Several mechanisms have been proposed: (1) The evaporation and condensation of silicate vapor, requiring gas heated to temperatures of $T\sim$ 1250--1450 K (probably at less than an AU from the parent star) \citep{Lodders2003,Gail2004,WoodenHarkerBrearley2005,Belletal1997};  (2) Annealing of amorphous silicate grains, requiring $T \gtrsim 1000$~K, (possibly occurring as far out as $\sim$2--3 AU in very hot, early disks) \citep{NuthJohnson2006,Woodenetal2007,Westphaletal2009};  and (3) Shock-heating and annealing in disk spiral arms. The first two of these are equilibrium processes, whereas the third relates to transient events that rely on the disk being massive enough to produce spiral arms.  It may, however, result in the production of crystalline silicates out as far as 10~AU \citep{HarkerDesch2002,ScottKrot2005,Woodenetal2007}\footnote{Note that although the strength of spiral shocks is expected to be greatest at many tens of AU \citep{ClarkeLodato2009}, the gas in these regions is too tenuous and cold to anneal silicate grains.}. These mechanisms have different chemical signatures.  Grain formation in long-lived, high-temperature regions of the disk is most likely to produce Mg-rich silicates due to the low oxygen fugacity expected to prevail in those regions.  Fe-bearing and Fe-rich crystalline silicates likely require a water-rich region of the disk in which to form, produced perhaps by migration and sublimation of icy bodies interior to the snow line, or in shocks in the outer disk \citep{Wooden2008}.  They may also require transient heating mechanisms to form without evaporating, unlike Mg-silicates, which can crystallize below short-time-scale--evaporation temperatures \citep{NuthJohnson2006}.  Pyroxene is thermodynamically favored over olivine when these minerals are formed by condensation, and, while annealing will tend to produce olivine from forsterite and pyroxene from enstatite, forsterite will convert to enstatite in long-lived ($\sim10^6$ years to completion), high-temperature conditions \citep{Gail2004,WoodenHarkerBrearley2005}.

Observationally, it has been known for some time that Oort Cloud comets contain crystalline-silicate material \citep{Hanneretal1994,Harkeretal2002,Hondaetal2004,WoodenWoodwardHarker2004}. These comets are believed to have formed primarily at distances of $\sim 5-10$~AU from the Sun, in a region of the disk that may have overlapped that where in situ crystalline silicate formation was possible. The high amounts of crystalline silicates recovered by {\em Stardust}, nonetheless, came as a surprise, since 81P/Wild~2 is a short period comet, that most probably formed in the outer disk, around the current orbits of Uranus and Neptune \citep{WoodenHarkerBrearley2005}. This is beyond any plausible source region for crystalline silicates. Subsequently, observations of the comet 9P/Tempel~1 have confirmed crystalline silicates to be an important component of that Jupiter-family comet as well \citep{Lisseetal2006}.  The compositional evidence from the {\em Stardust} samples points to diverse formation environments for the high temperature materials \citep{Wooden2008}. A predominance of Mg-rich silicate grains, together with the recovery of three calcium-aluminum (CAI-type) minerals that almost certainly formed by evaporation and condensation at $T > 1400$--2000~K, suggest a substantial contribution from the innermost, hottest, disk regions. However, some Fe-bearing and Fe-rich crystalline silicates were also recovered. There are also hints of igneous and aqueous alteration among some grains \citep{Joswiaketal2010,StodolnaJacobLeroux2010}, which, if confirmed, would place additional constraints on the timing of the outward transport and incorporation of these materials into the 81P/Wild~2 cometesimals. 

Direct comparison between the {\em Stardust} Solar System results and observations of other disks is difficult. Astronomical measurements are primarily sensitive to smaller orbital radii, and thus provide more important constraints on the degree and nature of processing than on radial transport. The observed mass ratio of crystalline-to-amorphous silicates varies greatly, both between disks \citep{vanBoekeletal2005,Watsonetal2009,Olofssonetal2009} and likely radially within a single disk \citep{vanBoekeletal2004,Olofssonetal2009}.  A study by \cite{Watsonetal2009} measured crystalline mass fractions in the inner disks ($\lesssim$10 AU) of 84 classical T~Tauri stars in the Taurus-Auriga star-forming region.  The measured mass fractions ranged from less than 0.5\% to more than 80\% despite the fact that the systems were all of similar ages, 1--2 Myr, and all were observed within a single star-forming cluster. A significant correlation is observed between the crystalline mass fraction and the extent of dust settling toward the disk midplane. This may be related to the link other studies have found between disk crystallinity and grain growth \citep{vanBoekeletal2005,Olofssonetal2009}, with characteristic crystalline grain sizes of a few microns.  No correlations were found relating to the mass or luminosity of the star, the mass of the disk, or the mass ratio of the star-disk system, all properties expected to affect heating and thermal processing within the disks.  These results agree with a previous study of Herbig Ae-star disks by \cite{vanBoekeletal2005} who found a correlation between disk crystallinity and stellar mass/luminosity that disappeared for $M_\star \lesssim 2.5M_\odot$.

In terms of composition, silicates within other protoplanetary disks show the same predominance of  Mg-rich grains as in the {\em Stardust} samples \citep{vanBoekeletal2004,vanBoekeletal2005,MolsterKemper2005, Watsonetal2009,Olofssonetal2009}. There are also marginally significant correlations between  crystallinity and stellar accretion rate. \cite{Watsonetal2009} found that for pyroxene, the trend in crystalline mass fraction was inverse to the mass-accretion rate onto the star, whereas for olivine, the trend was proportional to the accretion rate.   A study by \cite{Olofssonetal2009} found that disks of higher crystallinity tend to be dominated by enstatite (pyroxene-type) grains, and of lower crystallinity by forsterite (olivine-type) grains.  \cite{Olofssonetal2009} also suggest very heterogeneous mixing of silicate particles in disks, reporting a higher rate of crystalline-feature detection for the cold ($\lesssim$10 AU) than the warm ($\lesssim$1 AU) spectral features.  We will discuss later how some of these trends may be interpreted within the context of a turbulent transport model for grains within the disk.

\section{Methods}
\label{Disk}

The simplest model for studying the radial redistribution of particles within a disk assumes that the disk surface density is static and that the particles are small enough as to be perfectly coupled to the gas. Even in this limit, a range of other physical effects can be important and have been explored: mixing by (marginal) gravitational instability \citep{Boss2004,Boss2008}, the thermal evolution and production of annealed silicate grains \citep{Gail2001,BockeleeMorvanetal2002}, and vertical mixing and settling in an MHD model that includes dead-zone effects \citep{TurnerCarballidoSano2010}. Many of these studies agree that turbulent mixing can, in principle, be strong enough to explain the presence of crystalline silicates in comets.

Our goal in this work is to systematically incorporate a range of additional physical processes into what is still a relatively simple one-dimensional model for the evolution of particles within a gas disk. We focus on three effects:
\begin{itemize}
\item[(1)]
Imperfect coupling of particles to the gas. The largest particles captured by {\em Stardust} (about 20~$\mu$m) may -- depending upon the gas density -- be only marginally well-coupled.  Indeed, prior work that includes grain size and settling effects has found that large grains may settle out the the midplane and experience outward advection in a 2D stratified disk model \citep{Ciesla2007,Ciesla2009}.  Here we consider imperfect coupling to examine the feasibility of transporting large grain sizes radially outward into the more tenuous outer disk and comet-forming regions.
\item[(2)]
Disk evolution. The formation of material processed at high temperature almost certainly commenced early in the disk lifetime, when the disk would have been more massive, hotter, and more compact than the typical T~Tauri disk. As we will show later, the evolution of such disks can have a substantial impact on the radial transport of particles. This has been demonstrated explicitly by \cite{DullemondApaiWalch2006} and \cite{Dullemondetal2008}, who found that disks that form in initially more compact configurations produce greater outward mixing of hot material.
\item[(3)]
Uncertainties in the radial flow of gas at the midplane. Although the {\em vertically integrated} flow of gas in an active disk is assuredly inward at small orbital radii, the actual magnitude (and even direction) of flow at the midplane depends upon unknown aspects of the angular momentum transport within the disk. Indeed a flared viscous disk model with little to no vertical mixing yields an {\em outward} midplane flow \citep{Urpin1984,TakeuchiLin2002,Tscharnuteretal2009}, which could be very important for the transport of particles large enough to have partially settled \citep{KellerGail2004,Ciesla2007,Ciesla2009}.
\end{itemize}
We describe below our implementation of these effects within an otherwise standard one-dimensional model for the evolution of a gaseous protoplanetary disk, which is chosen so as to be consistent with observational estimates of the lifetime and accretion rate evolution of protoplanetary disks around low mass stars.

     \subsection{Disk Profile and Evolution}
     \label{Disk_Evolution}

We use the thin-disk model for the viscous evolution of the gas-disk surface density \citep{Pringle1981}.  We also include loss of disk gas due to photoevaporation by the central star, so that the surface-density profile evolves according to
\begin{eqnarray}
  \frac{\partial \Sigma_\mathrm{g}}{\partial t}
&=& 
  \frac{3}{R}
  \frac{\partial}{\partial R}
    \left[R^{1/2} \frac{\partial}{\partial R}
          \left(\nu \Sigma_\mathrm{g} R^{1/2}\right)
    \right]
\nonumber \\
&&- 
  \dot{\Sigma}_\mathrm{g,wind}\left(R,t\right)\,,
\label{sigma_evolv}
\end{eqnarray}
where $\Sigma_\mathrm{g}$ is the surface density, $t$ is time, $R$ is radial distance from the central star, and $\nu$ is the local disk viscosity. The $\dot{\Sigma}_\mathrm{g,wind}$-term is the loss due to photoevaporation by EUV \citep{Hollenbachetal1994} or X-rays \citep{Owenetal2009}.  Here we use the \cite{AlexanderArmitage2007} parameterization of the results of \citet{Fontetal2004}, which calculates photoevaporative mass loss due to a $10^{42}$~s$^{-1}$-photon flux of EUV from the central star. We do not here study the evolution of particles {\em during} photoevaporation (long after the presumed epoch when crystalline material must have been incorporated into comets or their progenitor bodies); we include photoevaporation only because it is essential for providing reasonable model-disk lifetimes.

We evolve the disk numerically using a standard time-explicit--finite-differencing scheme similar to that of \cite{PringleVerbuntWade1986}.  We use 600 grid cells spaced logarithmically.  Innermost and outermost grid points are placed at 0.1 and 15,000 AU, respectively.

We assume that the disk temperature and viscosity are both fixed and vertically uniform. We use a temperature profile of $T\left(R\right) = T_0 R^{-1/2}$, normalized to a temperature of 280~K at 1~AU. This is appropriate for a passive, flared disk after most of the gas infall has occurred and the initial rapid phase of accretion has slowed.  Note that very-early disk temperatures are expected to be much higher \citep{KenyonHartmann1987,Belletal1997,Tscharnuteretal2009}.  However, a more precise treatment of disk temperature would require models or assumptions involving, for example, disk chemistry and disk-forming infall rates.  For the disk viscosity, we use the $\alpha$-prescription of \cite{ShakuraSunyaev1973}, which gives
\begin{equation}
\nu\left(R\right) = \alpha c_\mathrm{s} H_\mathrm{g} \,,
\label{viscosity}
\end{equation}
where $\alpha$ is a non-dimensional scaling parameter, $c_\mathrm{s}$ is the local sound speed, and $H_\mathrm{g}$ is the local-disk scale height. Having assumed $T\propto R^{1/2}$, Equation (\ref{viscosity}) leads to $\nu\propto R$.  We adopt $\alpha = 10^{-2}$.  This value falls within the range of values estimated observationally \citep{Hartmannetal1998,HuesoGuillot2005,KingPringleLivio2007} and allows dissipation of our model disks in less than 10 Myr. The parameterization in Equation (\ref{viscosity}) assumes that the source of the viscosity is turbulence within the gas disk.  Equation (\ref{viscosity}) is also important for establishing the turbulent diffusion properties for the particle ensemble, as discussed in Section \ref{Transport_Diffusion}.

\begin{figure}
   \includegraphics[width=\columnwidth]{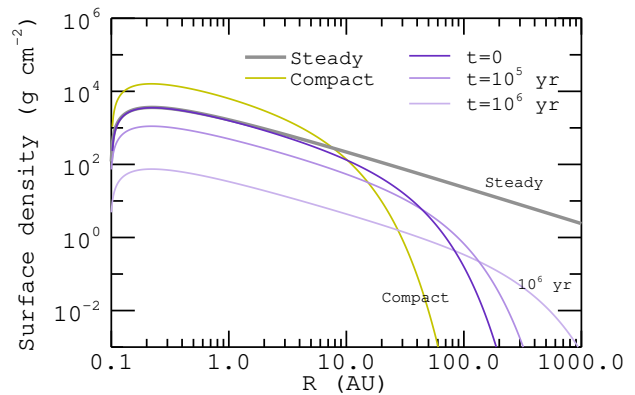}
   \caption{Surface-density profiles for the disk gas.  The purple curves track the time evolution of the nominal-disk model.  The yellow curve is the $t=0$ profile for the most compact ($R_\mathrm{d} = 5$ AU) disk model.  The thick grey curve is the profile of the static steady-disk model.}
   \label{sigmas}
   \end{figure}

The initial surface density distribution of the gas disk depends on the mass and angular momentum distribution of the parent cloud that collapsed to form the star-disk system \citep{LinPringle1990,HuesoGuillot2005}. Our evolving-disk models do not use a steady-disk profile, but rather assume a finite disk, so that at $t=0$, the surface density is, 
\begin{equation}
  \Sigma_\mathrm{g}\left( R, t=0 \right)
= 
  \frac{\dot{M}_0}{3 \pi \nu}
  \left( 1 - \sqrt{\frac{R_\mathrm{in}}{R}}\right)
  \exp{\left(-\frac{R}{R_\mathrm{d}}\right)} \,,
\label{sigma0}
\end{equation}
where $\dot{M}_0$ is the $t=0$ accretion rate onto the central star, $R_\mathrm{in}$ is the inner-disk boundary (set equal to the inner-grid boundary, $R_{1/2} = 0.099$ AU), and $R_\mathrm{d}$ is a variable controlling the compactness of the $t=0$ profile.  In our nominal-disk model, $R_\mathrm{d} = 20$ AU, and the initial disk mass of 0.03~M$_\odot$ dictates $\dot{M}_0 = 1.8 \times 10^{-7} \,M_\odot$ yr$^{-1}$. The evolution of this model is shown in Figure \ref{sigmas}. 
As the disk evolves, the surface density drops, and the disk spreads outward before dissipation via photoevaporation occurs after $\sim$5.5 Myr.  That disks may form in initially compact configurations has been discussed by several authors \citep{LinPringle1990,HuesoGuillot2005,DullemondNattaTesti2006} and may be indirectly supported by the steep disk surface densities calculated by \cite{Weidenschilling1977b} for estimates of the minimum-mass Solar nebula, and, more recently, by \cite{Desch2007}.  Depending on the strength of the disk viscosity, a compact initial configuration may also be necessary to process expected disk masses within observed disk lifetimes.

In Section \ref{Results_Compact} we vary $R_d$ (retaining $M_{\mathrm{d},0}=0.03 M_\odot$) to consider the effects of varied disk-forming conditions on the transport of particles.  To facilitate a comparison with other models, we also consider two models of dust motion within a static disk, including a steady-disk model where $R_\mathrm{d} = \infty$ (retaining $\dot{M}_0 = 1.8 \times 10^{-7} \,M_\odot$ yr$^{-1}$).  As shown in Figure \ref{sigmas}, this steady-disk model results in a disk profile that is matched to our $t=0$ nominal-disk profile in the inner-disk regions, but maintains a shallower profile and much larger gas surface densities in the outer disk.

     \subsection{Two Cases for Disk Gas Radial Velocity}
     \label{Disk_Velocity}

The mean radial flow of gas experienced by solid particles within the disk depends on aspects of disk turbulence that cannot yet be predicted from first principles. We maintain this uncertainty explicitly by considering two bounding cases for the gas flow, the ``accretion-flow" case and the ``midplane-flow" case. For the accretion-flow case, we use the vertically averaged gas velocity, $v_\mathrm{acc}$, \citep{Pringle1981}
\begin{equation}
  v_\mathrm{acc} 
= 
  - \frac{3}{\Sigma_\mathrm{g} R^{1/2}}
  \frac{\partial}{\partial R}\left(\nu \Sigma_\mathrm{g} R^{1/2}\right).
\label{vacc}
\end{equation}
This velocity can be derived from the 1D fluid equations and results in a gas flow that is predominantly inward. Importantly, however, there is a region of outward-flowing gas in the outer disk where the disk is expanding. The boundary of this region moves outward as the disk evolves and spreads.

In a real two-dimensional (2D) disk, the radial flow may vary with height above the disk midplane, and particles that settle toward the midplane may experience a very different radial-gas velocity than that predicted by Equation (\ref{vacc}). In particular, 2D $(R,z)$ disks in which the viscosity is given by a generalized $\alpha$ model with no vertical mixing of angular momentum yield an {\em outward} flow at the midplane at most disk radii \citep{Urpin1984,RozyczkaBodenheimerBell1994}. This flow is highly conducive to the transport of inner-disk grains to the outer disk.  For the midplane-flow case, we follow \cite{TakeuchiLin2002} and use the azimuthally symmetric Navier-Stokes equation to calculate $v_\mathrm{merid}$, the gas velocity at the midplane:
\begin{eqnarray}
  v_\mathrm{merid}  
  \frac{\partial}{\partial R}\left(R^2 \Omega_\mathrm{g}\right)
&=& 
  \frac{1}{R \rho_\mathrm{g}}\frac{\partial}{\partial R}
                             \left(R^3 \nu \rho_\mathrm{g} 
                                 \frac{\partial \Omega_\mathrm{g}}{\partial R}
                             \right)
\nonumber \\
       &+& \frac{R^2 \nu}{\rho_\mathrm{g}}\frac{\partial}{\partial z}
                                \left(\rho_\mathrm{g} 
                                      \frac{\partial \Omega_\mathrm{g}}
                                           {\partial z}
                                \right)
       - R^2 \frac{\partial \Omega_\mathrm{g}}{\partial t}
\label{vmerid}
\end{eqnarray}
Here, $\rho_\mathrm{g}$ is the local-gas volume density and $z$ is height above the midplane. $\Omega_\mathrm{g}$ is the local orbital angular velocity of the gas, determined via force balance in the radial direction \citep{TakeuchiLin2002},
\begin{equation}
  R\Omega_\mathrm{g}^2
- \frac{G M_\star R}{\left(R^2 + z^2\right)^{3/2}}
- \frac{1}{\rho_\mathrm{g}}\frac{\partial p}{\partial R}
= 
  0.
\label{omegag}
\end{equation}
In this equation, $M_\star$ is the mass of the central star (equal to 1 $M_\odot$), and $p$ is the local gas pressure.  Note that while \cite{TakeuchiLin2002} assumed a power law for the gas distribution and used that to derive a simplified expression for the flow structure of the gas, we have used the input of the disk surface density from our disk-evolution model to solve Equation (\ref{vmerid}) numerically by assuming vertically uniform temperature and viscosity.

\begin{figure}
   \includegraphics[width=\columnwidth]{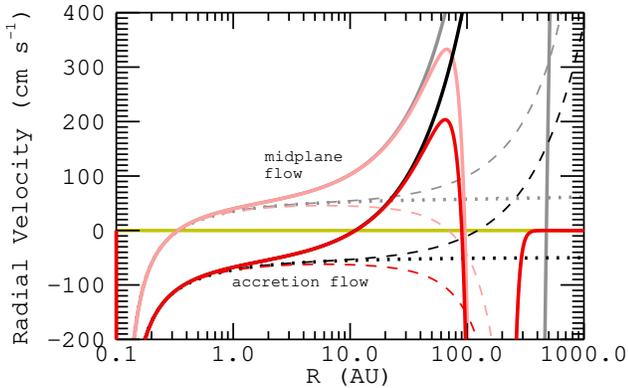}
   \caption{Radial advection velocities of the disk gas (black/grey) and 20 $\mu$m dust (red/pink). The black and red curves are for the accretion-flow case, and the paler-colored curves are for the midplane-flow case. Solid lines plot $t=0$, and dashed lines plot $t=1$ Myr, both for the nominal-disk model.  Dotted lines plot the gas velocity for the steady-disk model.}
   \label{velocities}
   \end{figure}

Figure~\ref{velocities} 
shows the two gas-velocity cases at $t=0$ and at $t = 1 \ {\rm Myr}$. Note that the outward expansion of the disk dominates both velocity cases at large disk radii.

\subsection{Particle transport}
\label{Transport}

Our model for particle transport tracks an ensemble of particles within a 1D gas-disk environment and subjects the particles to two non-gravitational forces: (1) aerodynamic drag against the mean gas flow and (2) turbulent diffusion of the ensemble within the disk gas. Coagulation is neglected. Our particle trajectories therefore consist of two components: (1) mean radial motion assuming Epstein drag against the mean-disk flow and (2) a random-walk component due to turbulent diffusion.

These two physical effects are represented numerically by two components to the dust  radial velocity: $v_{srd}$, the mean radial velocity induced by Epstein drag, and $v_\mathrm{turb}$, the turbulent random walk velocity. These velocities are calculated at each evolutionary time step of the gas disk model for each grid point of the model disk; values between grid points are linearly interpolated as necessary.  Together, these velocities are used to integrate the dust-grain trajectory as:
\begin{equation}
  r_d\left(t+\Delta t\right) 
= 
  r_d\left(t\right) + \Delta t \times \left(v_{srd} \pm v_\mathrm{turb}\right)
\,,\
\label{interpolate_traj}
\end{equation}
where $r_d$ is the dust-particle radial position in the disk, and $\Delta t$ is the time step. The time step is chosen locally, to insure accurate integration of particle trajectories, but is capped at the global disk evolution time step. The fidelity of this method in reproducing analytical results for trace-particle diffusion and transport in a static disk is discussed in the Appendix.  The relative motions of the dust and gas can vary from tightly coupled to divergent depending on the surface-area-to-mass ratio of the simulation particles and the local density of the disk gas.

     \subsection{Advection}
     \label{Transport_Advection}

We solve for the mean radial velocity of the dust grains, $v_{srd}$, subject to Epstein drag against the mean gas flow. In the Epstein-drag regime, the size of the particle is less than the mean free path of a gas molecule, and the drag scales with both the particle velocity and the thermal velocity of the gas as 
\citep{Weidenschilling1977a}:
\begin{equation}
  F_D 
= 
  m_\mathrm{d}\frac{C_R}{3}\rho_\mathrm{g}v_\mathrm{therm}
  \left(v_\mathrm{g}-v_\mathrm{d}\right)
\,,
\label{Epstein}
\end{equation}
where $C_R$ and $m_\mathrm{d}$ are the surface-area-to-mass ratio and mass of the dust grain, respectively, $\rho_\mathrm{g}$ is the local gas density, $v_\mathrm{therm} = \sqrt{8/\pi} c_\mathrm{s}$ is the local thermal velocity of the gas, and $v_\mathrm{g}$ and $v_\mathrm{d}$ are the gas and dust velocities.  In our model, we assume that $\rho_\mathrm{g} = \rho_\mathrm{g,mid} = \Sigma_\mathrm{g} / \sqrt{2\pi} H_\mathrm{g}$, the gas density at the disk midplane.  In the accretion-flow case, inputting the midplane density to Equation (\ref{Epstein}) is an approximation that selects for the tightest possible coupling between the gas and the particle ensemble at a given $R$; it leads to the most conservative estimate of outward mixing in this gas-flow case.  However, the midplane density is the most appropriate choice for use in the midplane-flow case, which presupposes that the entire particle ensemble has settled to the disk midplane.

To solve for the steady mean radial velocity of the dust grains, we must include both the drag force and the orbital motions in a polar coordinate system.  (See the appendix for the explicit force-balance equations of particle motion.)  We follow \cite{TakeuchiLin2002} and simplify these forces by assuming that the radial acceleration of the grain is zero, $dv_{r,\mathrm{d}} / dt \approx 0$, and that the azimuthal acceleration corresponds to a change in the Keplerian velocity with $R$, $dv_{\phi,\mathrm{d}} / dt \approx - v_\mathrm{K} v_{r,\mathrm{d}} / 2R$.  The equations for the radial and azimuthal motions of a grain now produce two equations for the steady mean radial velocity, $v_{srd}$, as a function of its steady azimuthal velocity, $v_{s\phi d}$:
\begin{eqnarray}
   v_{srd}
&=&
   v_{r,\mathrm{g}} + \frac{3\left(v_{s\phi d}^2-v_\mathrm{K}^2\right)}
                           {C_R R \rho_\mathrm{g} v_\mathrm{therm}}
\nonumber \\
   v_{srd}
&=&
 -\, \frac{1}{3}C_R\,R\rho_\mathrm{g}v_\mathrm{therm}
     \frac{\left(v_{s\phi d}-v_{\phi,\mathrm{g}}\right)}
          {\left(v_{s\phi d}-v_\mathrm{K}/2\right)} \,,
\label{dmotion2}
\end{eqnarray}
which we solve iteratively for $v_{srd}$ using the gas-disk bulk flow from our disk model at each radial grid point. In the Appendix, we demonstrate that the trajectories produced using the gridded-$v_{srd}$ values agree well with direct numeric integrations of the force-balance equations of grain motion for particles in disks with Epstein-drag stopping times ranging from less than 10$^{-4}$ Kepler times to greater than ten Kepler times.

Figure~\ref{velocities} shows how the mean radial velocities of the dust compare to the radial velocity of the gas. Very small grains are well coupled to the gas flow, but the crystalline silicates larger than 10 $\mu$m associated with the {\em Stardust} collection are large enough that at large distances from the star (tens of AU) and at late times in the disk evolution, the dust motion significantly departs from the gas flow.

     \subsection{Turbulent Diffusion}
     \label{Transport_Diffusion}

To model the turbulent diffusion of the particle ensemble, we add a random walk to the individual grain motion via the addition of a turbulent velocity component, $v_\mathrm{turb}$.  This turbulent velocity is set by the dust particle diffusivity, $D_\mathrm{p}$, and by the time step of the random walk motion; it is given by
\begin{equation}
v_\mathrm{turb} = \pm \sqrt{\frac{2D_\mathrm{p}}{\Delta t}} \,.
\label{vturb}
\end{equation}
In our particle-transport model, the time step of the random walk is limited by the time step of the gas-disk-evolution model, but is also allowed to be no greater than the local Kepler time, $1/\Omega_\mathrm{K}$ ($\Omega_\mathrm{K}$ is the Keplerian angular velocity), and no greater than either $\Delta R_5 / v_{srd}$ or $\Delta R_5 / v_\mathrm{turb}$, where $\Delta R_5$ is the radial distance across five grid cells in the direction of particle motion.

We follow \cite{YoudinLithwick2007} in calculating the radial diffusion coefficient of the dust grains via
\begin{equation}
  D_\mathrm{p} 
= 
  D_\mathrm{g} \, \frac{1+4\tau_\mathrm{s}^2}
                       {\left(1+\tau_\mathrm{s}^2\right)^2},
\label{diffusion}
\end{equation}
where $D_\mathrm{g}$ is the gas diffusion coefficient, and $\tau_\mathrm{s} = \Omega_\mathrm{K} t_\mathrm{stop}$.  Here, $t_\mathrm{stop}$ is the exponential-stopping time.  In the Epstein-drag regime, it is given by $t_\mathrm{stop} = 3 / \left(C_R \rho_\mathrm{g} v_\mathrm{therm}\right)$.  The dust and gas diffusion coefficients are equal throughout most of the disk, except in the outermost regions where the gas surface density is very low, and the dust diffusion coefficient drops to zero.  In our fiducial models, we assume that because the disk viscosity is derived from turbulence within the disk, $D_\mathrm{g} = \nu$.  In section \ref{Results_fSchmidt}, we vary $Sc=\nu/D_\mathrm{g}$ to evaluate the impact of this assumption on the potential for the outward mixing of inner-disk grains. 

While Equation (\ref{vturb}) provides an accurate representation of the particle diffusivity, it is not sufficient for producing accurate diffusion of the particle ensemble in our model gas disk.  Among other things, it does not account for the mass distribution of the disk gas.  In a real random walk, the time and distance of each step vary according to the local properties of the gas environment.  In our model, however, the time step is fixed by other, mostly numerical, considerations.  Therefore, to allow our particle ensemble to diffuse according to the actual gas distribution, we must calculate an appropriate probability for whether a given particle in the ensemble will step radially inward ($p_{in}$) or outward ($p_{out}$).

The proper weighting between $p_{in}$ and $p_{out}$ is a function of the imposed time step.  Instantaneously ($\Delta t \rightarrow 0$), we know that $p_{in}/p_{out}=1$, because $\Delta t=0$ represents only a single random walk encounter.  However, for $\Delta t > 0$, the probability that multiple encounters will occur within that time step becomes nonzero.  Furthermore, additional encounters will occur not at the initial position being considered, but either inward or outward of that location depending on the outcome of past encounters.  Therefore, the properties of the diffusive medium (the gas) outside the local point of interest play a role in determining $p_{in}/p_{out}$ for $\Delta t>0$.  The spatial extent of this region of interest is set by how far particles may travel for a given $\Delta t$.  For the particle ensemble, this is the root-mean-square of the displacement of the diffusing particles, or
\begin{equation}
   \Delta R
=
   \sqrt{\left<\left(\Delta x\right)^2\right>}
=
   v_\mathrm{turb}\Delta t
=
   \sqrt{ 2 D_\mathrm{p}\Delta t } \,\,.
\label{deltaR}
\end{equation}

The weighting for $p_{in}$ is set by the gas properties between $R-\Delta R$ and $R$; and for $p_{out}$ by the properties between $R$ and $R+\Delta R$. Asymptotically, the diffusion must yield everywhere a uniform particles(dust)-to-gas ratio.  Therefore, after an infinite time, the region with more gas mass must also have proportionally more dust particles.  However, because regions of higher diffusivity reach the steady state of uniform concentration more quickly, for a finite time step, proportionally more particles will also mix into a region of higher diffusivity than into one of lower diffusivity.  Therefore,
\begin{equation}
   \frac{p_\mathrm{in}}{p_\mathrm{out}}
=
   \frac{\left(M_\mathrm{g}\times D_\mathrm{p}\right)_{in}}
	{\left(M_\mathrm{g}\times D_\mathrm{p}\right)_{out}}
=
   \frac{\displaystyle \int_{R-\Delta R}^R D_\mathrm{p}\Sigma_\mathrm{g}
                                                R\,dR}
	{\displaystyle \int_R^{R+\Delta R} D_\mathrm{p}\Sigma_\mathrm{g} 
                                                R\,dR} \,.
\label{probincalc}
\end{equation}

To evaluate Equation (\ref{probincalc}) within our model gas disk, we assume uniform, mean values of the gas surface density and particle diffusivity between each grid point.  Because we calculate $p_\mathrm{in}$ once at each disk-evolution time step, but $\Delta t$ of the particle motions may vary locally for the individual grains, we calculate $p_\mathrm{in}$ at each grid point for both $\Delta t_\mathrm{evolve}$ and $\frac{1}{2}\Delta t_\mathrm{evolve}$, then interpolate parabolicly for other time-step sizes as needed, using $p_\mathrm{in}\left(\Delta t=0\right) = 0.5$.  

\begin{figure}
   \includegraphics[width=\columnwidth]{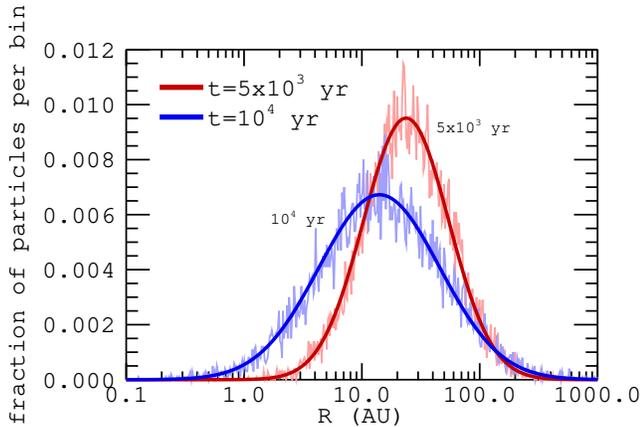}
   \caption{The fraction of particles per radial bin for our particle-transport simulations (pale lines) versus the analytical-expected values (dark lines). This figure corresponds to the \cite{ClarkePringle1988} example where $\nu \propto \Sigma_\mathrm{g}^{-1} \propto R^2$ in a static disk with an accretion-flow velocity, $v_\mathrm{acc} = -1.5 \nu / R$, and $\nu = 4.941\times 10^{14}$ cm$^2$ s$^{-1}$ at 1 AU.  The simulation was initiated at $t=0$ with 10$^4$ particles at 40 AU.}
   \label{aeqbeq2ex}
   \end{figure}

We have verified that this random-walk method does a good job of reproducing the expected diffusion profile of a contaminant within a 1D gas disk.  Figure~\ref{aeqbeq2ex} 
presents a comparison of the distribution of particles using our particle-transport model compared to the expected distribution for an analytical example derived by \cite{ClarkePringle1988}.  A more detailed discussion of the fidelity of our particle-diffusion model may be found in the Appendix.

\section{Effects of Individual Transport Processes}
\label{Overview}

To illustrate how each of the model components influences particle transport, we first consider a set of reduced models of successively greater complexity. We begin with a diffusion-only case for perfectly coupled particles, which we run keeping the gas disk profile fixed at its nominal $t=0$ form. We then add in the effects of radial gas velocity, particle size, and disk evolution. Unless otherwise stated, these simulations use 1000 particles initiated at 1 AU in the nominal-disk model.

\begin{figure}
   \includegraphics[width=\columnwidth]{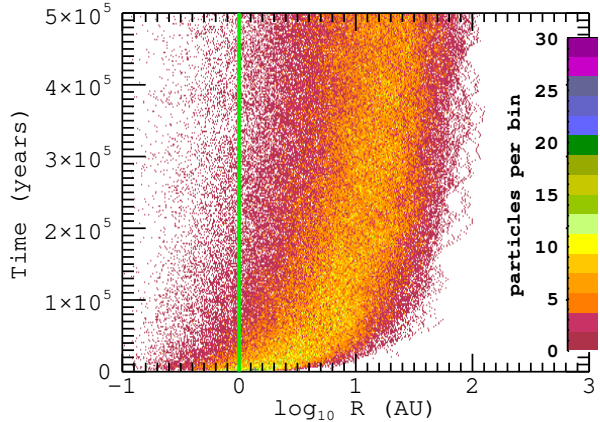}
   \caption{Basic diffusion: Map of the number of particles per grid space as a function of radial position and time.  We initiated 1000 well-coupled particles at 1 AU (denoted by the green line) in a static disk with zero radial-gas velocity.  Note that the radial width of the grid cells increases exponentially with distance from the host star.}
   \label{diffStaticWC0}
   \end{figure}

Figure \ref{diffStaticWC0} 
displays an image of the relative number of particles in each radial bin over time for the diffusion-only 
simulation\footnote{Note that the bin size increases exponentially with $R$, so this representation looks somewhat different from a plot of the particle surface density or concentration.}. 
Diffusion results in rapid (within $\sim 10^5$~yr) radial spreading over tens of AU, such that the particle distribution closely approximates that of the gas within a fraction of the disk lifetime. Some particles are also lost past the inner edge of the simulated disk. 

\begin{figure*}
   \includegraphics[width=\textwidth]{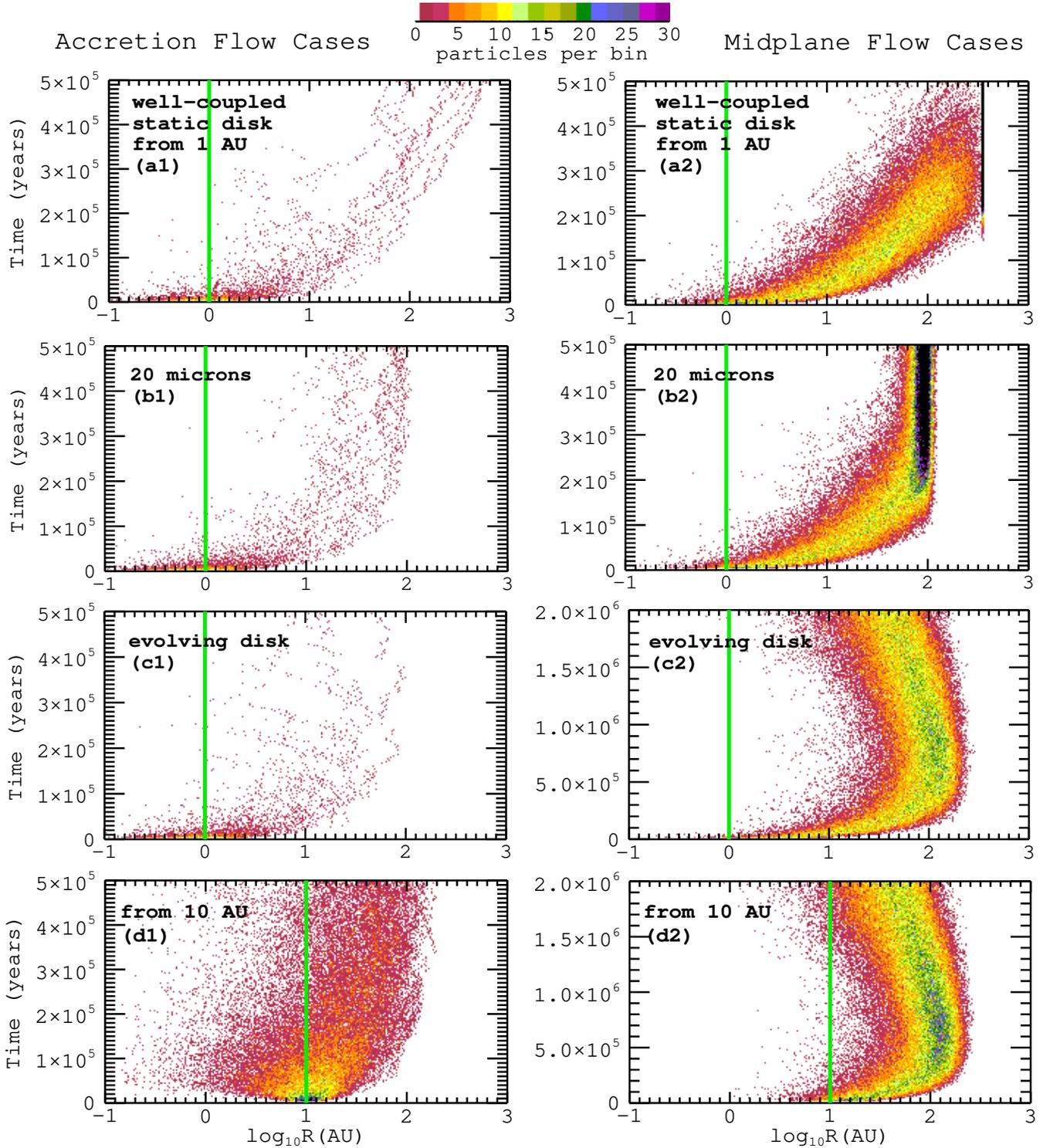}
   \caption{Diffusion scenarios: Maps of the diffusion of particles as in Figure 4  where the green line denotes starting location.  The right-side panels all include gas velocity in the accretion-flow case, and the left-side are for the midplane-flow case. The simulations in (b1) and (b2) add in the effects of non-zero particle size.  In (c1) and (c2) the disk surface density is also evolved with time, and the (d1), (d2) simulations are just like those above, but with the particles initiated at 10 AU.}
   \label{diffOverVpanels}
   \end{figure*}

Next, we include the effects of gas velocity by using the two radial-gas-velocity cases outlined in Section \ref{Disk_Velocity}.  The distribution maps for these simulations are shown in Figure \ref{diffOverVpanels} 
(a1, a2), and here the gross effects of mostly inward vs mostly outward gas velocities are clear.  In the accretion-flow case, where most of the gas flow is inward, the majority of the particles are quickly lost onto the central star. Only $\sim$2\% of the particles remain in the simulation after $10^4$ years. However, the particles that do remain in the disk reach the outer-disk region and continue to move outward, because the steep gas distribution in the outer disk causes the accretion flow there to be outward.  In the midplane-flow case, less than 10\% of the particles are lost. Particles are instead swept rapidly outward with the gas flow and become trapped at the edge of the static disk-gas distribution, where the value of the disk surface density falls rapidly toward zero (at about 340 AU).

After turbulent diffusion and gas velocity, we add in the effects of particle size.  The results of simulations using 20 $\mu$m particles are shown in panels (b1) and (b2) of Figure \ref{diffOverVpanels} . The effects of particle size are most clearly visible in the midplane-flow case where particles are still advected outward as before, but pile up at smaller radial distances (around 90 AU) than in the simulations with the well-coupled particles. What is happening here is that beyond $\sim$115 AU, the disk gas is too diffuse to support the outward advection of particles of this size.  Instead, headwind drag dominates particle motion, pushing particles inward.  In the accretion-flow case, the effects of a non-negligible particle size are similar.  Though the dominant behavior of particle motion here remains the loss of most of the particles inward onto the parent star, those 20 $\mu$m particles that do reach the region of outward accretion flow are also confined to smaller radii than their massless counterparts.

Finally, we include the effects of disk evolution.  As the disk evolves, the surface density drops.  Panels (c1) and (c2) of Figure \ref{diffOverVpanels} show that in the accretion-flow case, this drop in density results in a sustained inward loss of the particles; in the midplane-flow case, the majority of particles are still retained and advected outward.  In fact, some particles reach greater distances than in the static-disk model, because, as the disk expands, the surface density no longer drops off as steeply around 100 AU.  However, after about $t=0.5$ Myr, the decrease in disk surface density begins to dominate over the effect of the disk expansion.  The gas can no longer support 20 $\mu$m particles at such large distances, and the particle distribution begins to move inward.

So far, we have presented simulations of particles initiated quite close to the parent star ($R=1$ AU).  In panels (d1) and (d2) of Figure \ref{diffOverVpanels} we see that if the particles are initiated further out in the disk, the particle motions are roughly the same, with the major exception that significantly more particles are retained in the accretion-flow case.  For particles initiated at 10 AU in this case, about 60\% remain in the disk after $10^5$ years.  In the midplane-flow case, particles initiated at 10 AU reach the furthest regions of the disk slightly faster, but otherwise behave the same as when initiated at 1 AU.  Therefore, while both gas-flow cases can turbulently diffuse particles over wide regions of the disk, the accretion-flow case selectively retains particles initiated at larger radii and rapidly loses particles initiated at smaller radii.

To summarize, the basics of particle mixing in these transport simulations are that:
\begin{enumerate}
\item Turbulent diffusion rapidly spreads the particle distribution over many tens of AU.
\item Where advection tends toward the loss of particles inward onto the parent star, there will be a preferential rapid depletion in grains initiated close to the parent star.
\item Even in the case of outward-flowing disk gas (the midplane-flow case), small dust grains may still spiral inward at radii (or epochs) where the disk is locally tenuous.
\end{enumerate}
All of these effects will be reflected in the availability of inner-disk particles in the comet-forming regions.

\section{Results}
\label{Results}

In this section, we study how the parameters of the system affect the efficiency of mixing inner-disk particles out to large distances.  These parameters include the structure/direction of the radial gas flow, the sizes of the dust particles, the diffusivity relative to the disk viscosity, and the initial distribution and evolution of the disk gas mass. We vary these parameters relative to a fiducial model that assumes 20 $\mu$m particles (all particle sizes assume a dust-particle internal density of 3 g cm$^{-3}$), a Schmidt number of $Sc=1$ ($D_\mathrm{g}=\nu$), and an evolving nominal-disk model. The nominal-disk begins with a total gas mass of 0.03 M$_\odot$ and an exponential fall-off radius of $R_d = 20$ AU, so that at $t=0$, 61\% of the disk mass is concentrated within 20 AU of the star.  

In each model, the particles (10,000 for the accretion-flow runs, 2,000 for the midplane-flow) are initialized with a uniform radial distribution between 0.5 and 10 AU. Although this is not an exact match to the initial gas distribution at these radii, it allows us to fully sample the likely source regions for crystalline silicates within the disk. We are interested, primarily, in the question of what fraction of these particles reach the ``comet-forming region" (defined, for this paper, as the disk beyond 25~AU) at any time during the evolution of the disk. However, in some physical models, crystalline silicate formation is confined to smaller radial extents.  Therefore, we also consider a subset of our simulation particles divided into three smaller source regions: the inner-quarter region (0.5--2.5 AU, 2106 particles in the accretion-flow simulations, 421 particles in the midplane-flow simulations), the inner-half region (0.5--5 AU, 4737 and 947 particles), and the outer-half region (5--10 AU, 5263 and 1053 particles).

     \subsection{Transport in Different Radial Gas-Flow Cases}
     \label{Results_Flow}

\begin{figure}
   \includegraphics[width=\columnwidth]{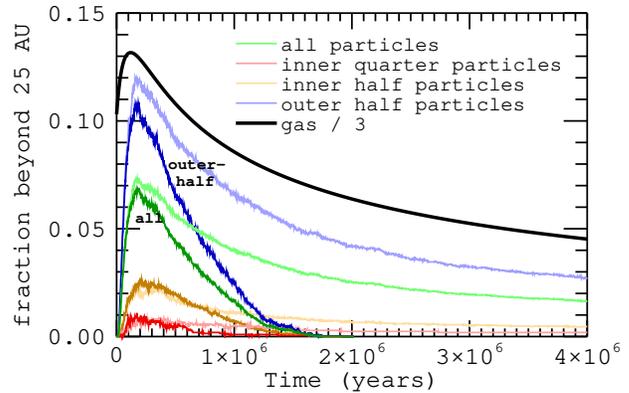}
   \caption{Fraction of simulation particles beyond 25 AU for different source regions compared to the fraction of the total disk gas mass currently beyond 25 AU (divided by three).  Pale-colored lines denote particles well-coupled to the gas; dark colors denote 20 $\mu$m-sized particles.  The inner-quarter source region is 0.5--2.5 AU; the inner-half source region is 0.5--5 AU; the outer-half source region is 5--10 AU.  Simulation run in the nominal disk model using the accretion-flow case.}
   \label{vacc_by25}
   \end{figure}

Figure~\ref{vacc_by25} 
shows the baseline results for particle mixing if the effective gas velocity seen by the particles is that of the accretion-flow case. We plot the fraction of 20~$\mu$m particles beyond 25~AU that originate from each of the source regions, together with the same quantity for perfectly coupled particles (i.e. of negligible size). We also show the {\em gas} fraction beyond 25~AU. As expected from the reduced models, diffusion is fast enough to mix some particles upstream into the comet-forming region. Indeed, the first particles pass 25~AU in less than 20,000 years.  However, loss of particles by accretion onto the star is also rapid, especially close to the inner edge of the disk. After $10^5$ years only $\sim$30\% of all 20~$\mu$m particles remain in the disk, and most of those originated in the outer half of our total source distribution (the peak fractions of particles in the comet-forming region are 2.6\% from the inner-half zone, but 11\% from the outer-half zone). The lifetime of particles that do manage to attain large radii is also limited, first by the fact that the transition radius that divides gas inflow from outflow itself moves out as the disk evolves, and second because declining gas densities in the outer disk eventually preclude retention of 20~$\mu$m particles. These effects mean that by $t=10^6$ years, less than 1\% of 20 $\mu$m particles exist in the disk beyond 25 AU, even if they started in the outer-half source region.  By $t=2.1$ Myr, all have been lost inward onto the parent star.

\begin{figure}
   \includegraphics[width=\columnwidth]{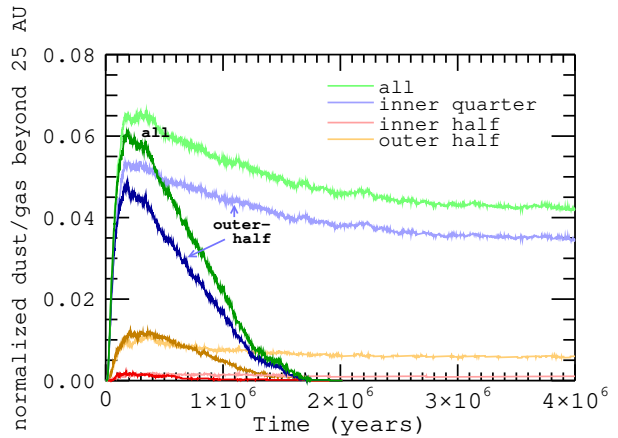}
   \caption{Normalized concentration of particles beyond 25 AU. [See Equation (\ref{normalized_concentration}).]  The pale-colored lines denote well-coupled particles.  The dark lines are for 20 $\mu$m particles. Nominal disk model.  Accretion-flow case.}
   \label{vacc_cby25}
   \end{figure}

To quantify the effect of these mixing patterns on the abundance of hot, inner-disk grains in the comet-forming regions, we plot in Figure~\ref{vacc_cby25} 
the normalized concentration, $C_N$, of source-region particles beyond 25~AU as a function of time.  We define the normalized concentration as the number of particles in a region divided by the gas mass in that region, then normalized by the initial number of source-region particles divided by the $t=0$ gas mass in that source region:
\begin{equation}
  C_N
=
  \frac{\left(\mathrm{\# particles}\right)_\mathrm{outer \, disk} / 
                  \left(\mathrm{gas \, mass}\right)_\mathrm{outer \, disk}}
          {\left(\mathrm{\# particles}\right)_{\mathrm{source},\,t=0} / 
                  \left(\mathrm{gas \, mass}\right)_{\mathrm{source},\,t=0}}
\label{normalized_concentration}
\end{equation}
With this definition, $C_N$ is like a scaled dust-to-gas ratio, where $C_N=1$ means that the ratio of particles-of-interest--to--gas in the region of interest (the outer disk) is the same as where those particles originated at the time that they formed. We note that while the denominator of Equation~(\ref{normalized_concentration}) is almost constant across our source region (i.e. the initial dust to gas ratio is a fixed value to within 10\%), the normalized concentration beyond 25~AU is an average over a broad region. The local values can vary widely, depending upon the parameters of the disk.  For the accretion-flow case simulations, the highest local values of the normalized concentration typically lie just outside of 25 AU, with a lower local $C_N$ in the regions hundreds of AU from the parent star.

From Figure~\ref{vacc_cby25}, we see that the peak values of the normalized concentration -- and thus the maximum extent to which the outer disk can be contaminated by particles from the hot inner regions -- is relatively modest. For 20~$\mu$m particles, the peak in $C_N$ from the all source region is approximately 6\%. This is actually higher than the value obtained from the outer half source region, reflecting the fact that although a greater fraction of particles mix outward from this region, there is less mass there to start with. We also observe that particle size strongly influences the lifetime of particles in the outer disk. For well-coupled particles, $C_N$ declines only modestly from its peak at a few hundred thousand years out to several Myr. This reflects the fact that well-coupled particles that survive the initial evolution have thoroughly mixed with most of the entire disk gas mass, so that their subsequent loss is largely in proportion with the disk-gas accretion rate. By contrast, 20~$\mu$m particles are lost much faster, and only significantly contaminate the outer disk between $\sim 5\times10^4$ and $8\times10^5$ years. 

\begin{figure}
   \includegraphics[width=\columnwidth]{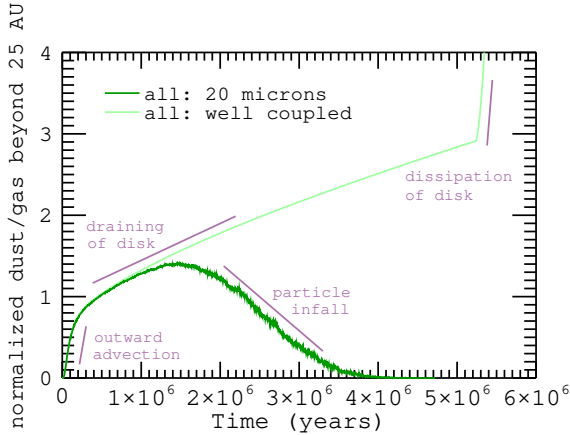}
   \caption{Midplane-flow case: Normalized concentration of simulation particles beyond 25 AU for all particles initiated between 0.5 and 10 AU.  Nominal disk model.}
   \label{vmerid_cby25}
   \end{figure}

Dramatically higher efficiencies of particle retention are obtained if, instead of experiencing the mean gas flow, the particles settle and experience an outward flow at the disk midplane (recall that this requires little or no vertical mixing of angular momentum, so that the steep radial density gradient at the midplane results in outflow). Results for this midplane-flow case are plotted in Figure~\ref{vmerid_cby25}.  
In this limit, few 20~$\mu$m particles are initially lost inward onto the parent star. 98\% remain in the disk after $10^5$ years, and around 90\% of particles from each source region are transported to beyond 25 AU, with only a slight selection against particles from the innermost regions of the disk. This means that the normalized concentration curves for the different source regions are all nearly the same, only scaled relative to a given source region's innate ability to contaminate the outer disk - the relative fraction of $t=0$ disk mass in that source region.

Because 20 $\mu$m and well-coupled particles are swept to large distances and remain there for a long time, there is a period when the change of normalized concentration beyond 25 AU is a function solely of the loss of disk gas mass via accretion.  Without the concurrent loss of accreting particles, the particle-to-gas ratio reaches values even higher than in the original 0.5--10 AU source regions.  This is particularly so for the well-coupled particles, which never leave the outer disk once they have entered it.  The normalized concentration of 20 $\mu$m particles drops to half-maximum at around $t\sim 2.6$ Myr, about three times the contamination time-window given in the accretion-flow simulations.

Note, however, that our model cannot account for the radial mixing that would occur in the presence of vertical mixing and a vertically stratified gas flow.  If the gas flow is rapidly inward far from the midplane and outward near the midplane, then vertical mixing will lead to radial mixing, even in the absence of radial diffusion.  \cite{Ciesla2007,Ciesla2009} discusses in detail how a given degree of vertical mixing can lead to enhanced mixing of grain populations or to the partial segregation of inward-flowing and outward flowing populations.  Therefore, the normalized concentration for the midplane-flow case merely represents the extreme upper limit on the extent and duration of the contamination of the outer disk by inner-disk material.

The results imply that the efficacy of the fiducial disk model ($R_d = 20$~AU, ${\rm Sc} = 1$) for outward transport of 20~$\mu$m grains depends upon both the location (and extent) of the the source region for crystalline silicates, and upon the radial gas flow experienced by the particles. In the accretion flow case, the peak of contamination of the outer disk (beyond 25~AU) would result in a crystalline-to-amorphous ratio of $\sim$6\%. Attaining this limit requires that at $t=0$ all silicate grains in the entire inner disk out to 10 AU were crystalline (and likewise assumes that all beyond that were amorphous). If, instead, the source of crystalline silicates is confined to inward of 2.5~AU then the peak $C_N$ beyond 25 AU is only about 0.2\%. Moreover, we find that even these modest degrees of contamination are relatively short-lived. 20 $\mu$m-sized grains (matched to the upper-size end of the {\em Stardust} grains) only have a contamination lifetime in the outer disk of $\lesssim$~1 Myr. Unless the timing of the epoch when grains became assembled into cometary material happened to coincide with the peak of the contamination, the crystalline fraction would be diluted, either by grains grown locally in the cold comet-forming region or by those raining inward from the even-colder outer regions of the disk. Finally, our results for the accretion flow case might be yet further reduced by the fact that some crystalline silicate formation mechanisms may not yield 100\% crystalline fractions out to large radii. For example, if the primary source of crystalline silicates is shocks out to $\sim$10~AU, but a sizable fraction of silicates in that region remain amorphous, then that cuts into the potential maximum in the crystalline-to-amorphous ratio produced in the outer disk.

In the midplane-flow case, on the other hand, substantial outward transport can occur almost irrespective of the location of the source region.  If there is an outward flow of gas at the disk midplane, then grains that are sufficiently well-settled will readily reach large distances \citep{Ciesla2009}.  An outward-midplane flow also allows material that has been transported outward to persist there for a longer period of time.  Our midplane-flow simulations produce an upper-limit maximum $C_N$ in the outer disk of 30\% for the inner-quarter source region.  Even if vertical mixing means that only a fraction of the maximum can be achieved, it is still plausible to believe that the CAI-like grains captured during the {\em Stardust} mission may have reached the outer disk via mixing within the disk gas alone.   This scenario may also be consistent with the observations of \cite{Watsonetal2009}, who find higher crystalline-to-amorphous silicate mass ratios in disks where the grains appear more settled toward the disk midplane, and of \cite{Olofssonetal2009}, who find that possibly a large fraction of crystalline silicates exist in the outer regions ($\lesssim$10 AU versus $\lesssim$1 AU) of some disks.

     \subsection{Transport of Differently Sized Particles}
     \label{Results_Size}

\begin{figure}
   \includegraphics[width=\columnwidth]{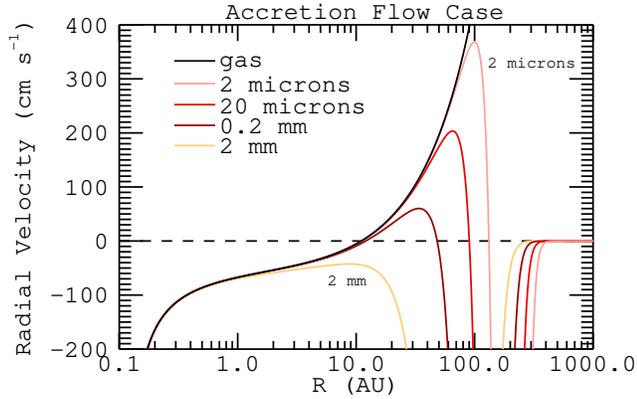}
   \caption{The mean-radial velocity of particles of different sizes in the nominal disk model at $t=0$, accretion-flow case. Sizes listed assume an internal-particle density of 3 g cm$^{-3}$.}
   \label{sizesVels}
   \end{figure}

In Section \ref{Overview}, we demonstrated that non-negligible particle size bars particles from the outermost regions of the disk, because at large distances, the gas is so tenuous that the particles decouple from the gas motions, and inward-pointing head-wind drag dominates.  Figure \ref{sizesVels} 
plots the dust-mean-radial velocities for several different particle sizes in the nominal-disk model at $t=0$ for the accretion-flow case.  (In the midplane-flow case, the curves are similar, only shifted toward outward velocities as in Figure \ref{velocities}.) Though turbulent diffusion can send particles upstream of their average radial-velocity flow, grains are effectively barred from the regions of the disk beyond where their average-radial velocity falls to large, inward-pointing values.  It is clear from Figure \ref{sizesVels} that the range within the disk that dust grains may occupy depends strongly on the particle size, and that the mixing achievable out to the comet-forming regions will become sharply limited as one approaches millimeter particle sizes.  Also, where this cutoff occurs changes, not only in time as the disk thins, but also from disk to disk depending on the total mass and mass distribution of each system.

\begin{figure}
   \includegraphics[width=\columnwidth]{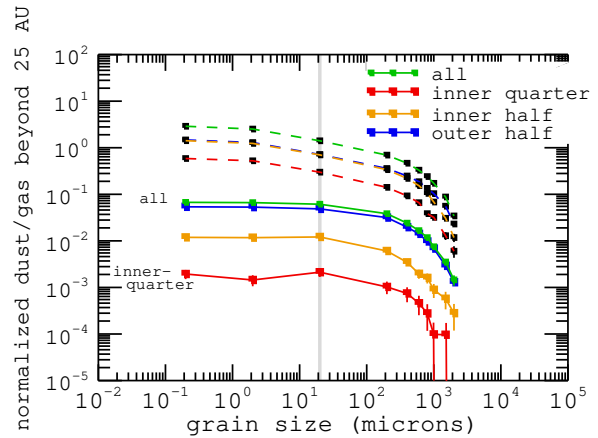}
   \caption{Maximum values of normalized concentrations beyond 25 AU as a function of grain size.  Solid lines denote simulations run in the accretion-flow case; dashed lines with black markers denote the midplane-flow case.  Error-bars mark the $\sqrt{N}$ noise of the peak values.  Nominal disk model. Grey line marks the baseline case of 20 $\mu$m-sized particles. In the midplane-flow case for the smallest particles, peak values shown are for $t=5.2$ Myr, just prior to photoevaporative dissipation of the disk.}
   \label{vgrainsize_cby25max}
   \end{figure}

We parameterize the outward mixing of grains as a function of size by plotting in Figure \ref{vgrainsize_cby25max} 
the peak values of the normalized concentration of grains beyond 25 AU as a function of grain size.  The error bars show statistical errors based on the number of particles beyond 25~AU in each run. We plot the maximum values of $C_N$ from each of the different source regions, as well as the upper-limit values provided by the midplane-flow simulations (dashed lines with black markers). While the magnitude of outward mixing varies by orders of magnitude across these populations, the trends with grain size are the same across all sets: as seen previously, the highest normalized concentrations of inner-disk particles in the outer disk occur for the smallest grain sizes, up to a few tens of microns, and the peak-$C_N$ values drop off sharply at around millimeter grain sizes. The results of \cite{Ciesla2007,Ciesla2009} for a vertically stratified disk flow suggest that the loss of larger grain sizes may be partially mitigated by settling toward the midplane, thereby increasing the relative outward transport and retention of these particles.  Still, the concurrent drop in our midplane-flow upper-limit concentrations suggests that this can only hold back the loss of large grains from the outer disk to a point.  For 2 mm-sized particles, for example, even the midplane-flow runs yield a peak normalized concentration beyond 25~AU for all simulation particles of only $\sim$3.4\%.

\begin{figure}
   \includegraphics[width=\columnwidth]{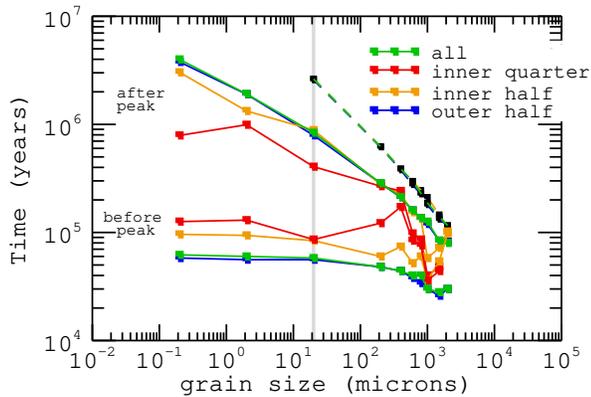}
   \caption{Time-windows for the peak values of normalized concentrations beyond 25 AU as a function of grain size and for different source regions.  The lower set of solid curves mark the latest pre-peak time of half-maximum concentration in the accretion-flow case.  The upper curves mark the earliest post-peak time of half maximum concentration for the accretion-flow case (solid lines) and the midplane-flow case (dashed lines with black markers). Nominal disk model. In the midplane-flow case, the smallest particles remain in the outer disk longest and never drop below half-maximum concentration beyond 25 AU.}
   \label{vgrainsize_timewindow}
   \end{figure}

The other important variable to consider is the timing of these mixing events.  In Figure \ref{vgrainsize_timewindow}, 
we depict, as a function of particle size, the time-frame over which the normalized concentration of inner-disk particles in the comet-forming region is at a value of half-maximum or greater.  For each source region (for the accretion-flow case simulations) we plot the latest time of half-maximum concentration prior to the peak and the earliest time of half-maximum after the peak. (Due to limited particle statistics, some of these quantities are quite noisy.) We also include the boundary of post-peak half-maximum concentration in the midplane-flow simulations to illustrate when headwind-drag inspiral will overwhelm other physical effects even in the most extreme gas-flow scenario.

\begin{figure}
   \includegraphics[width=\columnwidth]{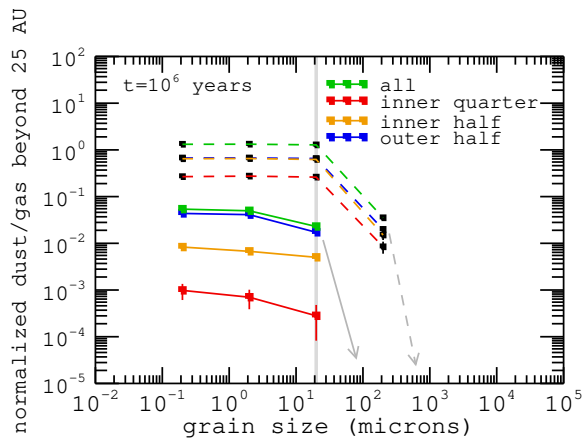}
   \caption{Normalized concentration beyond 25 AU at $t=1$ Myr as a function of grain size. The largest particles are all depleted from beyond 25 AU by 1 Myr. Dashed lines with black markers denote the midplane-flow simulations. Nominal disk model.}
   \label{vgrainsize_cby6}
   \end{figure}

Figure \ref{vgrainsize_timewindow} demonstrates that the smallest inner-disk particles remain mixed into the outer disk for millions of years, and that even inner-disk grains as large as 20 $\mu$m may be noticeably present at large distances for $\sim$1--2 Myr.  However, while the mixing of all grains outward is relatively rapid in these simulations, not only do the larger particles experience relatively less outward mixing, but that mixing is similarly short-lived.  Even in the midplane-flow simulations, post-peak half-maximum concentration is reached by $t\sim 6.2\times 10^5$ years for 0.2 mm particles and as soon as $t\sim 1.1\times 10^5$ years for 2 mm particles.  The fact that larger inner-disk particles are short-lived in the outer disk is demonstrated more bluntly in Figure \ref{vgrainsize_cby6}, 
where we plot the normalized concentration as a function of grain size at $t=1$ Myr.  While the mixing effects are relatively uniform for small particles at this time, no particles 0.2 mm and larger remain in the outer disk in the accretion-flow simulation, and the upper-limits of the midplane-flow simulations suggest that even in that extreme those large-grain populations are in rapid decline, if not already disappeared.

These results imply that it is the upper envelope of particle sizes recovered by {\em Stardust} that provides the most stringent constraints on disk evolution and particle transport. In our simulations, particles at least as large as 2 $\mu$m appear to behave as though well-mixed with the disk gas.  Therefore, the presence of high-temperature particles in this size range within a Jupiter-family comet can be largely explained by outward mixing of disk material aided by some disk expansion from an initially compact state. However, the particles recovered by {\em Stardust} include crystalline silicates as large as 20 $\mu$m.  While the outward transport of 20 $\mu$m particles to the comet-forming region does occur in our models, these particles do not remain well coupled to the disk gas indefinitely.  This suggests that either (1) mixing of particles and planetesimal formation in the Solar nebula occurred in a disk massive enough for 20 $\mu$m particles to remain well coupled to the gas motions, or that (2) the formation of comets or cometesimals occurred within the first million years or so of disk evolution.   This second constraint may be in agreement with {\em Stardust} results.   Early studies found no aqueous minerals in the {\em Stardust} materials, suggesting that the outward transport and incorporation of these materials into the 81P/Wild~2 cometesimals occurred very early, before the primary accretion and fragmentation of the chondritic-asteroid parent bodies \citep{Wooden2008}. However, the possible identification of igneous materials in the {\em Stardust} samples may contradict this scenario or else put interesting time-line constraints on the formation time-scales of planetesimals and cometesimals across the range of Solar System radii \citep{StodolnaJacobLeroux2010,Joswiaketal2010}. Another constraint linked closely with time is the large-end grain size for which our model predicts measurable outward mixing.  Not only do the peak (and midplane-flow--upper-limit peak) values of $C_N$ beyond 25 AU drop off markedly for grains a few millimeters in size, but the time-frame for that outward transport is also restricted (to barely more than 10$^5$ years for 2 mm-sized grains), so that comet-formation would have to be quite rapid  to capture such large inner-disk grains if they did appear in the comet-forming regions.

The transport of different-sized particles is also of interest for models that include the effects of grain growth.  Some observational studies of disks suggest a link between disk crystallinity and grain growth \citep{vanBoekeletal2005,Olofssonetal2009}, with characteristic crystalline grain sizes of a few microns (distinctly larger than typical ISM grains).  At the very least, grain growth and crystallization likely occur on similar time scales. The results presented here suggest that as dust grains grow, they should become less likely to enter and more likely to leave the outer regions of the disk.  Large grains become mostly confined to small disk radii where the disk gas is still dense enough to support them.  We know that growth at centimeter and meter scales poses a constraint to disk models and planetesimal formation, because particles of this size are expected to fall inward toward the parent star very rapidly, depriving the disk of these solids.  However, growth to millimeter scales may also place constraints on cometesimal formation if the outer regions of the disk cannot support particles of even these sizes. At the very least, it could restrict the composition of bodies formed in the outer disk, if most of the particles available are those falling inward from large AU.

     \subsection{Transport Varying the Diffusivity}
     \label{Results_fSchmidt}

\begin{figure}
   \includegraphics[width=\columnwidth]{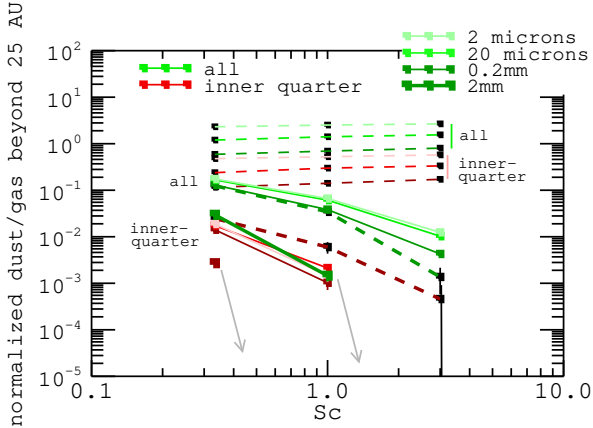}
   \caption{Maximum values of normalized concentration beyond 25 AU as a function of Schmidt number for a range of grain sizes and two source regions. Low diffusivity (large $Sc$) tends to restrict particles from reaching the outer disk. Dashed lines with black markers denote the midplane-flow simulations. Nominal disk model.}
   \label{vfsch_cby25max}
   \end{figure}

Next, we explore how the outward mixing of inner-disk grains depends on the Schmidt number, which is the ratio of the disk viscosity to the gaseous diffusivity (so that a lower Schmidt number means more relative diffusivity.) In \cite{PavlyuchenkovDullemond2007} the authors argue for an $Sc$ lower-limit of $1/3$.  Therefore, we have run simulations of particle transport varying the Schmidt number by a factor of three, and in Figure \ref{vfsch_cby25max} 
we plot the values of maximum normalized concentration beyond 25 AU for four particle sizes and two source regions as a function of $Sc$.  The primary effect shown in Figure \ref{vfsch_cby25max} is that a higher diffusivity can lead to a substantially higher degree of outward transport in the general accretion-flow case of our particle-transport simulations.  Here we see variations of an order-of-magnitude or more in the peak outer-disk normalized concentrations between the highest ($Sc=1/3$) and lowest ($Sc=3$) diffusivity simulations across the range of particle sizes considered.  This is less than the variation predicted in the model scenario of \cite{PavlyuchenkovDullemond2007} ($\sigma_{Sc=1/3} / \sigma_{Sc=3} \propto \left(R/R_\mathrm{source}\right)^4$), but is still quite a substantial effect considering that our normalized concentrations are reported for the whole of the outer disk and that our pool of contaminants is limited to those present in the source regions at $t=0$ only.

More generally though, the Schmidt number controls the relative importance of advection versus diffusion of particles.  When the diffusivity is low, advection dominates, so that more grains reach the outer disk if the bulk flow is outward, and fewer do so if their bulk flow is inward (including the case of 2 mm particles in the outward-flowing--gas midplane-flow case whose radial advection is dominated by headwind drag).  Conversely, high diffusivity pushes the system in the direction of a flow-independent state where diffusion, both within the disk gas distribution and inward onto the parent star, is the dominant term. The diffusivity can also have an important effect on outer-disk contamination from the inner-most source regions.  For 20 $\mu$m-sized particles from the inner-quarter source region, $Sc=1/3$ simulations produce a peak normalized concentration in the comet-forming region of 1.7\%, up from 0.2\% in the baseline $Sc=1$ simulations.  In the low-diffusivity simulations inner-quarter source grains of all sizes are completely absent from the comet-forming region unless directly advected there by outward-flowing gas in the midplane-flow simulations.

\begin{figure}
   \includegraphics[width=\columnwidth]{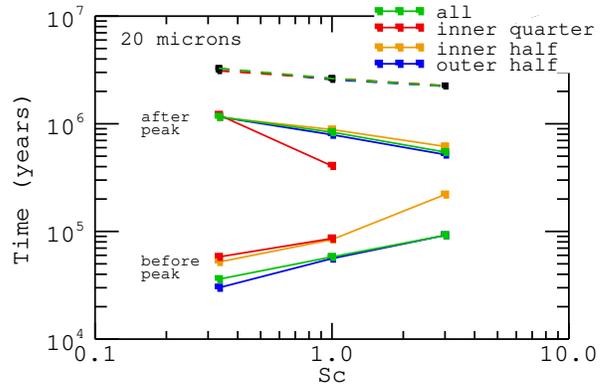}
   \caption{Time-windows for the peak values of normalized concentration of 20 $\mu$m particles beyond 25 AU as a function of Schmidt number ($Sc \equiv \nu /D_\mathrm{g}$).  Lower solid curves mark the latest pre-peak time of half-maximum concentration in the accretion-flow simulations.  Upper curves mark the earliest post-peak time of half-maximum concentration for the accretion-flow (solid lines) and midplane-flow (dashed lines) simulations. Nominal disk model.}
   \label{vfsch_timewindow}
   \end{figure}

Finally, to some degree, the diffusivity also affects the timing of outward mixing.  This is plotted for 20 $\mu$m-sized particles in Figure \ref{vfsch_timewindow}.  
The higher diffusivity simulations allow higher normalized concentrations in the outer disk to occur sooner and to last longer ($\sim 4\times 10^4$--$1.1\times 10^6$ years half-max to half-max), whereas for lower diffusivity the outer-disk normalized concentration peaks later and more briefly ($\sim 9\times 10^4$--$5.5\times 10^5$ years).

Our results for the variation in transport efficiency with Schmidt number suggest that two distinctly different disk models could be consistent with observations. One possibility is that the diffusivity is high, as suggested by \cite{PavlyuchenkovDullemond2007}. In this case, it seems plausible that a significant fraction of grains, formed in high temperature regions close to the star, can end up in the comet-forming region, even if the disk is highly turbulent and no significant settling occurs. The recovery of CAI-type grains by {\em Stardust} -- whose condensation requires especially high temperatures -- poses the strongest constraint on this scenario, and we have not demonstrated explicitly that it is possible. However, given that the early disk was certainly hotter than the disk which we have modeled, it is plausible that enough material from the innermost disk could reach large enough radii. Hot outflows near the disk midplane, which \cite{Tscharnuteretal2009} find very early in the star/disk formation process, would assist outward transport from the innermost disk.

A qualitatively distinct disk model with low diffusivity is also conceivable. Our results suggest that if ${\rm Sc}$ is significantly larger than unity, there is negligible transport of crystalline silicates subject to the accretion flow to the region where Jupiter family comets form. This is true even if silicates can form at radii as large as 10~AU. A low diffusivity disk, however, could potentially be favorable to the establishment of an outward midplane flow, which {\em is} able to move settled particles outward efficiently. \cite{Ciesla2009} showed that a transport scenario dominated by high-altitude--inward and low-altitude--outward advection can lead to segregated grain populations with different processing histories. This is possibly compatible with disk observations of \cite{Olofssonetal2009} that suggest some disks are more crystalline at larger distances.

     \subsection{Transport in More/Less Compact Disks}
     \label{Results_Compact}

Finally, we assess the role of the disk mass distribution and its evolution on outward transport. First, we contrast the results of runs assuming a static disk with those of the evolving disk, and then move on to exploring the effects of the initial compactness of the disk in the evolving-disk model.  Of our two static-disk models, the first uses the $t=0$ profile of the nominal-disk held static (the static0-disk model).  The second is the steady-disk limit of the $t=0$ nominal-disk profile, where the disk surface density follows Equation (\ref{sigma0}), with $R_d = \infty$ (the steady-disk model). 

\begin{figure}
   \includegraphics[width=\columnwidth]{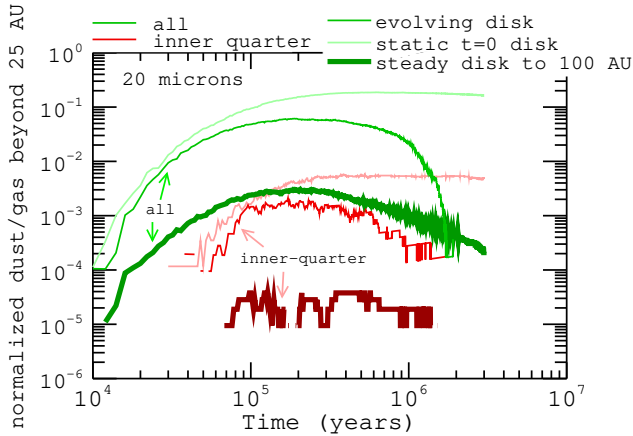}
   \caption{For the static disk models, the normalized concentration of 20 $\mu$m particles beyond 25 AU for two source regions.  Pale-curves denote the static0 disk model.  Dark, heavy curves denote the steady disk model assuming a gas mass truncated beyond 100 AU.  The curves for the nominal, evolving disk model are included for comparison.  Simulations run in the accretion-flow case.}
   \label{statics_cby25}
   \end{figure}

Results from the two static-disk models are shown in Figure \ref{statics_cby25}, 
which plots the normalized concentration of 20 $\mu$m particles beyond 25 AU with time for two different source regions.  The mass per AU of the gas disk does not drop off in the outer disk for the steady-disk model but is instead continuous out to the edge of our simulation space.  Therefore, the normalized concentrations beyond 25 AU for this disk model are calculated using the disk gas mass only out to 100 AU (for a total 0--100 AU disk mass of 0.162 $M_\odot$ --- 0.03 $M_\odot$ of gas is contained within the first 21 AU in the steady-disk model).  Figure~\ref{statics_cby25} includes the results for the evolving nominal-disk model for comparison.   The most obvious result from Figure \ref{statics_cby25} is that the two static-disk models, together with the nominal evolving-disk model, form a hierarchy in outward mixing that is based on disk structure.  In the static0-disk model the region of outward-flowing accretion flow remains fixed relatively close to the central star, and those outward velocities are fairly rapid.  Therefore, more than twice the peak fraction of nominal-disk particles move to the outer disk (and stay there, as per the static-disk-model tendencies seen in Section \ref{Overview}).   The surface-density structure in the steady-disk model means that in the accretion-flow case the gas radial-velocity is everywhere inward, and hence the number of 20 $\mu$m particles beyond 25 AU peaks at less than 40\% of the value for the corresponding nominal evolving-disk simulations, and those particles eventually all fall back inward onto the parent star, --- and more rapidly than in the evolving disk simulations. Roughly 23\% remain in the disk after $10^5$ years (when 30\% remain in the nominal-disk accretion-flow simulations).

However, in terms of the normalized concentration of inner-disk particles in the outer disk, the disk-gas mass distribution also really sets the steady-disk simulations apart from the rest.  With so much more mass in the outer disk than the other two models, the steady-disk model is that much less intrinsically capable of contaminating the outer disk with inner-disk grains than are the disks with a more compact disk mass distribution.  Finally, for the upper limits placed by the midplane-flow simulations both static-disk models send more than 90\% of the simulation particles from all source regions to beyond 25 AU; the outward advection is only more rapid in the static0-disk model.  Therefore, the upper bounds on the mixing in these models are set only by their relative source-region and comet-forming region disk-gas mass ratios.

Although some trends are the same for static as compared to evolving disk models, the results imply that the evolution of the disk -- particularly at early times -- has a substantial impact on particle transport. Steady-disk models tend to underestimate the degree of outward particle transport possible in protoplanetary disks in the early stages of disk evolution, when the most processing of high-temperature materials occurs. The outward transport of particles is substantially more efficient at early times, consistent with the results of \cite{BockeleeMorvanetal2002} and \cite{Ciesla2009}, even here assuming a fixed diffusivity. How strong the effects of evolution are will depend, of course, on the details of the initial conditions for the disk. To examine this dependence, we vary the initial compactness of our evolving-disk model (parameterized by $R_d$ in Equation~\ref{sigma0}).  We consider $R_\mathrm{d} =$ 5, 10, and 40 AU, in contrast to $R_\mathrm{d} =$ 20 AU in the nominal-disk model.  We retain 0.03 $M_\odot$ as the starting mass of the disk, so that the lifetimes of the model disks vary between 3.6--6.9 Myr, and the $t=0$ mass accretion rates between $8.3\times 10^{-7}$--$8.6\times 10^{-8} M_\odot$ yr$^{-1}$. We do not consider possible variations in heating for disks forming in more or less compact configurations, retaining the static temperature profile described in Section~\ref{Disk_Evolution}\footnote{We also continue to employ the same uniform radial distribution to initiate our simulation particles. One should note, however, that the gas-mass--per-AU is no longer roughly uniform across the source regions in the $R_d=10$ and 5 AU disks. For $R_d=10$ AU, the ratio of the average-gas-mass--per-AU for the inner-quarter source region to the outer-half source region is $\sim 1.5$, and for $R_d=5$ AU it is $\sim 2.6$. This affects the accuracy of the normalized concentration values that we report for these disks (especially for the widest source regions), but it does not alter the trends with $R_d$ repeated across source-region populations.}. 

\begin{figure}
   \includegraphics[width=\columnwidth]{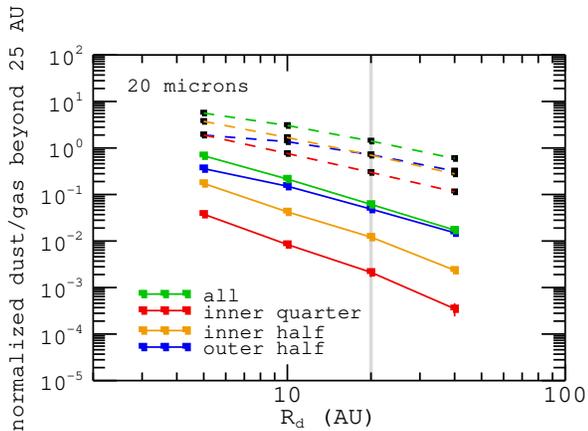}
   \caption{Maximum values for the normalized concentration of 20 $\mu$m particles beyond 25 AU as a function of initial disk compactness. Dashed lines with black markers denote the midplane-flow simulations.  Grey line marks the baseline case of $R_d=20$ AU. Note that the most compact disks both send a larger fraction of simulation particles beyond 25 AU and have a larger fraction of disk material in the source regions at $t=0$.}
   \label{vRd_cby25max_20mic}
   \end{figure}

Reducing the compactness of the model disk has two consequences: the disk initially expands more rapidly, and it does so from an initial state where a greater fraction of the mass is at small radii (and thus hot and potentially able to form crystalline material). The combined magnitude of these effects is shown in Figure~\ref{vRd_cby25max_20mic}, 
where we plot the maximum normalized concentration of 20 $\mu$m particles beyond 25 AU as a function of $R_d$. Variations of approximately one order of magnitude in $R_d$ result in nearly two orders of magnitude difference in the peak concentrations for each source region. For the smallest disk -- with $R_d = 5$~AU -- the normalized concentration of all simulation particles peaks at $\sim$68\%, more than an order of magnitude in excess of the value for the nominal simulations. A substantial part of this increase is due to the larger reservoir of mass that a compact disk has at small radii, which can potentially contaminate the (proportionally less massive) outer disk.  That this mass-distribution effect is important is made clear by the fact that even the upper-limit concentrations from the midplane-flow simulations show a strong (order-of-magnitude) trend with $R_d$.

\begin{figure}
   \includegraphics[width=\columnwidth]{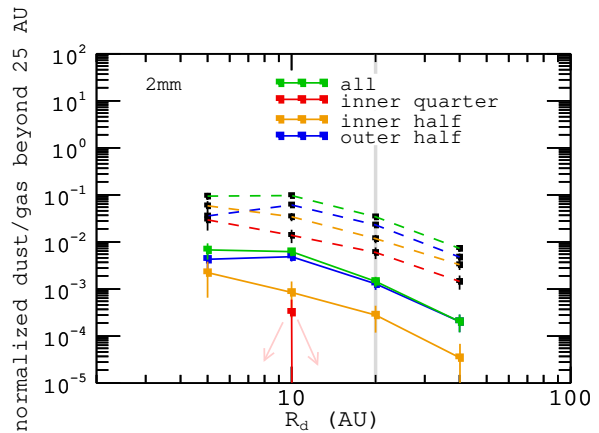}
   \caption{Maximum values for the normalized concentration of 2mm particles beyond 25 AU as a function of initial disk compactness. Dashed lines with black markers denote the midplane-flow simulations.  In the accretion-flow case, almost no 2mm particles from the inner-quarter source region reach beyond 25 AU.}
   \label{vRd_cby25max_2mm}
   \end{figure}

The fact that the outer disks of the initially more-compact-disk models are less massive (and more tenuous) does place some constraints on the outward mixing of large grains in our particle-transport simulations.  In Figure \ref{vRd_cby25max_2mm} 
we plot the peak normalized concentration of 2 mm particles beyond 25 AU as a  function of $R_d$.  All of the disks, of course, have less outward transport of these larger grains, but in our simulations, the peak in the fraction of particles sent to beyond 25 AU occurs for $R_d=10$ AU, which sends up to 0.25\% of all 2 mm-simulation particles beyond 25 AU (4\% in the midplane-flow case) compared to less than 0.1\% for $R_d=5$ AU (less than 2\% in the midplane-flow case).  Because of the larger source region mass for $R_d=5$ AU, this drop in fractional outward transport constitutes a leveling off in the maximum-achievable normalized concentrations of inner-disk particles in the outer disk.

\begin{figure}
   \includegraphics[width=\columnwidth]{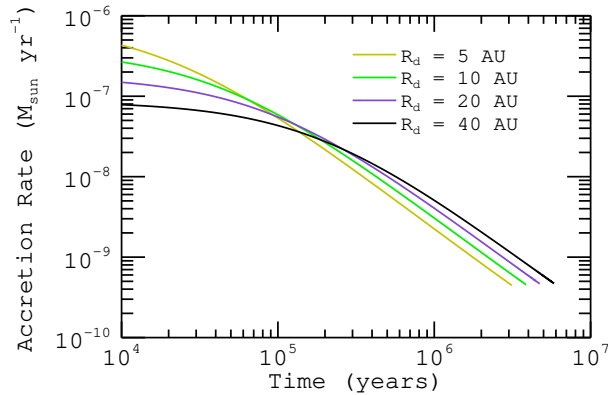}
   \caption{Mass accretion rates for disk models initiated with the same total disk mass, but with different initial exponential cutoff radii ($R_d$). Note the reversal in ordering from early to late times. All disks were initiated at $t=0$ with a total mass of 0.03 $M_\odot$.}
   \label{compactMacc}
   \end{figure}

Comparing these results to observations of crystalline silicates in other disks allows us to construct a possible scenario for the formation and transport of high-temperature minerals in those disks.  In the \cite{Watsonetal2009} survey, the disks observed were around T~Tauri stars $\sim$1--2 Myr old and showed a slight correlation between the crystalline-silicate mass fraction and the measured accretion rate onto the star. As shown in Figure \ref{compactMacc} 
for our model disks, $t=1$~Myr corresponds to a time when those disks that were initially most compact (with high $t=0$ accretion rates) now have lower accretion rates than the initially more-extended disks.  Therefore, small grains that formed at small AU near $t=0$ should be more broadly distributed in disks that now have lower accretion rates.  However, lower accretion rates also mean disks that are more tenuous.  The initially most compact disks lose mass the fastest, causing their outward transport efficiency to drop rapidly with time.  Therefore, grains that formed at small AU after $t\sim 2\times10^5$ yr may be better distributed in the initially more-extended disks, which now have higher accretion rates.

These properties could potentially explain the opposing trends between crystallinity and observed accretion rate reported by \cite{Watsonetal2009}. Pyroxene is favored over olivine when the minerals are formed by condensation \citep{Gail2004,WoodenHarkerBrearley2005}. Therefore, a possible scenario is that pyroxene formed primarily at very early times when the disks were hottest and became most broadly distributed throughout the initially most-compact disks. Most of the olivine, on the other hand, may have formed over a longer period of time.  It might, therefore, have become more broadly distributed within disks that retained significant surface densities for longer times (the initially more-extended disks).  This hypothesis does not, however, explain why the \cite{Watsonetal2009} survey found no correlation between crystalline mass fraction and disk mass.  

\section{Conclusions}
\label{Conclusions}

Observations of comets \citep{Hanneretal1994,WoodenWoodwardHarker2004} and sample return from the {\em Stardust} mission \citep{Brownleeetal2006} suggest that outward transport of high-temperature solids to the comet-forming regions is a necessary component of any successful model of the early Solar nebula.  We have constructed a 1D model of viscous disk evolution and particle transport that we use to study the relationships between the evolving disk mass distribution and the local conditions that are supportive of the outward mixing of grains, specifically including self-consistent treatment of the dust grains' surface-area-to-mass ratios in our transport terms.  We have examined patterns of outward mixing considering variations in the radial gas-flow structure, the sizes of dust grains, the model's Schmidt number, and the initial compactness of the model disk's gas mass distribution.

We find that a range of disk models are able to account for the presence of high-temperature--inner-disk particles in the comet-forming region.  As in \cite{DullemondApaiWalch2006}, the most favorable models involve rapid early expansion of initially quite compact disks. These conditions can result in inner-disk material flowing to the outer disk on a short time scale. In agreement with \cite{Ciesla2009}, we find that outward transport is more efficient at early times due largely to stronger gas outflows. Even more favorable are cases involving outward-flowing gas at the disk midplane \citep{KellerGail2004,Ciesla2007,Ciesla2009}, which are capable of delivering substantial fractions of grains inward of 1~AU to the outer regions of the disk. It is not known whether such outward flows exist in real disks (or, whether the particles are sufficiently settled to experience them consistently). Whether such midplane flows are {\em required} to explain the {\em Stardust} results depends upon the Schmidt number, which at its lower limit (highest relative diffusivity) could allow for a relatively high degree of general outward mixing, even from relatively near the parent star. While the mean accretion flow is less efficient at transporting particles outward, significant outward flows of crystalline material would be possible if the disk were initially both compact and able to form silicates out to a distance of several AU.

Our results suggest that it is the largest particles recovered by {\em Stardust} that place the tightest constraints on the evolution of the early Solar nebula. Particles as large 2 $\mu$m are well coupled to the disk gas, and plausible levels of turbulent diffusivity allow them to reach the outer disk from small radii in a wide variety of models. Our simulations show that 20~$\mu$m particles can also reach the outer disk --- consistent with the {\em Stardust} results --- as long as they were incorporated into larger bodies within the first 1--2 Myr of the disk evolution. For even larger particles, those of a few mm or larger, none of our disk models admit significant outward transport to the comet-forming region, regardless of the proposed source region. Even grains of a few hundreds of microns in size should be rare, since their residence time in the outer disk is very limited. Discovery of such grains would, within the context of the model developed here, require a more massive disk during the time of their formation, transport, and incorporation into cometesimals.

The existence of substantial uncertainties in the disk physics currently precludes an observational determination of the characteristic radii at which crystalline silicate formation occurs. The strongest constraints come from observations of CAI-like grains, which probably formed at fractions of an AU from the Sun. Only a subset of our accretion-flow simulations are likely compatible with this finding. A broader range of models in which the midplane gas flow is outward are viable, since these models retain particles in the disk for a longer period of time and facilitate the co-existence of grain populations of markedly different processing histories. Generically, we expect that provided sufficient outward transport is possible, turbulent diffusion will cause particles to become radially well-mixed over tens of AU so that incorporation of CAI-like grains and Fe-rich crystalline silicates (possibly formed in water-rich shocks out to 10~AU) into the same body is plausible.

For astronomical observations,  our results suggest that the mass and size of the disk at an early epoch are critical parameters in determining the viability and extent of outward particle transport. These disk properties are inherited from the mass and angular-momentum structure of the cloud that collapses to form the star, and would vary from system to system. We expect this to result in variations in the crystalline fraction at large radii in disks around different stars. We have also noted that the time scale during which conditions allow substantial outward transport is only a small fraction of the disk lifetime.  A limited time for outward mixing is both interesting and puzzling when paired with the emerging  evidence for aqueous and/or igneous grains among the {\em Stardust} samples \citep{Wooden2008,Joswiaketal2010,StodolnaJacobLeroux2010}. However, it may allow the {\em Stardust} results to be reconciled with other observations that require compositional gradients within the disk to be maintained. A possible scenario is that an early stage of rapid expansion drove particles into the outer disk where they were rapidly locked up into larger icy bodies. Once the disk had expanded, the gas flow became more uniform, limiting the outward transport of solids.

\acknowledgements

We thank the referee for his or her thorough reports, and in particular for alerting us to inconsistencies in an earlier version of our turbulent transport model.
This work was supported by NASA's Origins of Solar Systems program (NNX09AB90G) by NASA's Astrophysics Theory and Fundamental Physics program (NNX07AH08G), and by the NSF's Division of Astronomical Sciences (0807471).

\appendix
\section{VERIFYING THE FIDELITY OF THE PARTICLE TRANSPORT MODEL}
\label{Fidelity}

Our particle-transport model consists of two pieces.  The first is the simple mean-radial velocity of particles immersed in a radial and azimuthal gas flow.  It is represented by a calculation of the mean-radial velocity at each grid point in the model disk at each time step.  The second piece is the turbulent diffusion of the particle ensemble.  It is represented as a random walk with respect to the background mean-radial velocity.  For the random walk, we calculate a turbulent velocity for each particle at each time step and the probability that the particle will randomly walk inward versus outward.  Using these velocities and random-walk probabilities, we can then do a simple integration of the radial motion of each particle using Equation (\ref{interpolate_traj}).

In this appendix, we demonstrate that this simplified approach to particle mixing and transport does, in fact, accurately represent the physics necessary to our particle-transport model.  In Section \ref{Fidelity_Advection}, we demonstrate that we have chosen appropriate local velocities for the dust-mean-radial velocities and that those velocities are sufficient to represent radial particle motion.  We compare our radial-dust trajectories to trajectories produced using direct numerical integration of the the dust-particle equations of motion, Equations~(\ref{dmotionaAA}) and (\ref{dmotionbAA}).  In Section \ref{Fidelity_Diffusion}, we compare test-case simulations from our particle-transport model to analytical solutions of the diffusion of a contaminate in an accreting disk given by \cite{ClarkePringle1988}.  We demonstrate that the collective behavior of our random-walking particles produces the equivalent of diffusion in a gas disk.

     \subsection{Advection Compared to Force-Balance Integration}
     \label{Fidelity_Advection}

Our particle-transport simulations use an approximation technique to calculate particle trajectories.  Here we compare the trajectories produced using this technique to precision numerical integrations of those trajectories.  The equations of motions for a particle in a laminar disk-gas flow in the Epstein-drag regime are \citep{TakeuchiLin2002}
\begin{eqnarray}
  \frac{d v_{r,\mathrm{d}}}{dt}
&=& 
  \frac{v_{\phi,\mathrm{d}}^2-v_\mathrm{K}^2}{R} 
- \frac{1}{3}C_R\,\rho_\mathrm{g} v_{\mathrm{therm}}
  \left(v_{r,\mathrm{d}}-v_{r,\mathrm{g}}\right) \,,
\label{dmotionaAA}
\\
  \frac{dv_{\phi,\mathrm{d}}}{dt}
&=& 
  -\frac{v_{\phi,\mathrm{d}}v_{r,\mathrm{d}}}{R} 
- \frac{1}{3}C_R\,\rho_\mathrm{g} v_\mathrm{therm}
  \left(v_{\phi,\mathrm{d}}-v_{\phi,\mathrm{g}}\right) \,,
\label{dmotionbAA}
\end{eqnarray}
where $v_{r,\mathrm{d}}$ and $v_{\phi,\mathrm{d}}$ are the radial and azimuthal velocities of the dust particle; $v_{r,\mathrm{g}}$ and $v_{\phi,\mathrm{g}}$ are the local radial and azimuthal velocities of the disk gas; $v_\mathrm{K}$ is the local Keplerian velocity; $C_R$ is the particle surface-area-to-mass ratio; and $\rho_\mathrm{g}$ and $v_\mathrm{therm}$ are the local-gas volume density and thermal velocity, respectively.

In the case of laminar-gas flow with no turbulent forcing of the particle motions, we can numerically solve for each particle trajectory directly.  To do this, we use a Runge-Kutta-style integrator, where Equations~(\ref{dmotionaAA}) and (\ref{dmotionbAA}) are coupled with
\begin{equation}
\frac{dr_\mathrm{d}}{dt} = v_{r,\mathrm{d}}
\label{rungakuttaeq}
\end{equation}
to define the particle motion radially within the disk.  The integrator uses step-size adjustment to reach a relative accuracy of $10^{-5}$--$10^{-6}$ in $r_\mathrm{d}$, $v_{r,\mathrm{d}}$, and $v_{\phi,\mathrm{d}}$.  Note, however, that the integrator becomes numerically expensive when the individual terms in Equations~(\ref{dmotionaAA}) and (\ref{dmotionbAA}) are large but sum to values that are small.  Therefore, if the initially computed acceleration ($dv_{r,\mathrm{d}}/dt$ or $dv_{\phi,\mathrm{d}}/dt$ or both) is very small ($<10^{-14}$ cm s$^{-2}$), the integrator sets the small acceleration equal to zero.  Also, if $v_{\phi,\mathrm{d}}\sim v_\mathrm{K}$ (10$^{-10}$ relative difference) then the $v^2_{\phi,\mathrm{d}}-v_\mathrm{K}^2$-term in Equation (\ref{dmotionaAA}) is set equal to zero, but only if $\displaystyle \frac{1}{3}C_R\rho_\mathrm{g}v_\mathrm{therm} \geq \frac{v_\mathrm{K}}{R}$.  If $v_{\phi,\mathrm{d}}\sim v_\mathrm{K}$ but the second condition is not met, then the $v_{\phi,\mathrm{d}}^2-v_\mathrm{K}^2$-term is still important and the integrator instead calculates the particle trajectory to the lower relative accuracy of $10^{-5}$.

\renewcommand{\thefigure}{A\arabic{figure}}
\setcounter{figure}{0}

\begin{figure}
   \includegraphics[width=\textwidth]{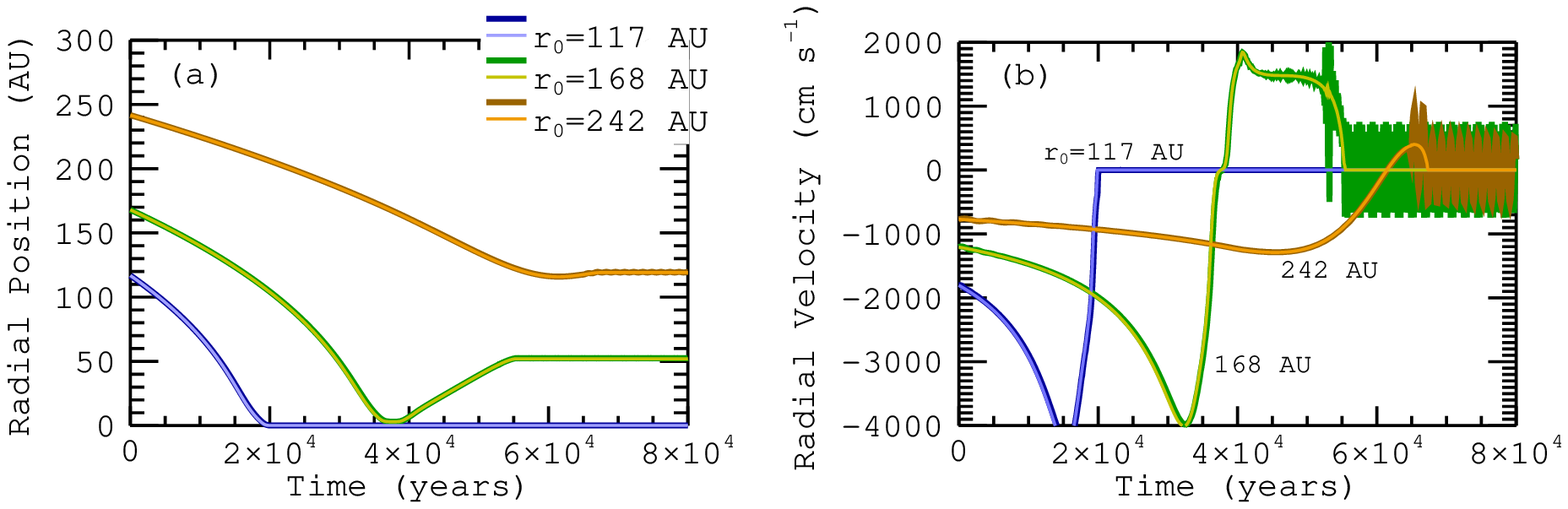}
   \caption{Radial position (a) and radial velocity (b) of simulated particle trajectories.  Comparison between precision numerical integrations (heavy, dark lines) and the particle-transport-model trajectories (thin, pale lines).  In this figure, 2 mm-sized particles were initiated at three different radial positions within a tenuous, evolving disk and were subject to the midplane-flow gas-velocity case.}
   \label{mvsAA2007hsm2}
   \end{figure}

Most of the comparison simulations presented below are run using the midplane-flow-velocity case, as defined in Section \ref{Disk_Velocity} of this paper. We begin by presenting particle trajectories calculated within an initially tenuous disk ($R_d=500$~AU, $\Sigma_\mathrm{g}\left(1\mathrm{AU}\right)\sim 1$~g cm$^{-2}$ at $t=0$) that is photoevaporatively cleared within $\lesssim$10$^{5}$ years.  Radial particle trajectories in such disks can be complex and so provide good tests of our particle-transport model.  Figure~\ref{mvsAA2007hsm2} 
plots, for both numeric integrations and particle-transport simulations, the trajectories and radial velocities of 0.2 mm-sized particles initiated at three different radial locations within this tenuous disk.  Here, such large particles are only loosely coupled to the gas flow, with exponential-stopping times ($T_s$) at $t=0$ of a few to greater than ten Kepler times.  Nevertheless, there is good agreement between our particle-transport simulations and the more precise numeric integrations.  

\begin{figure}
   \includegraphics[width=\textwidth]{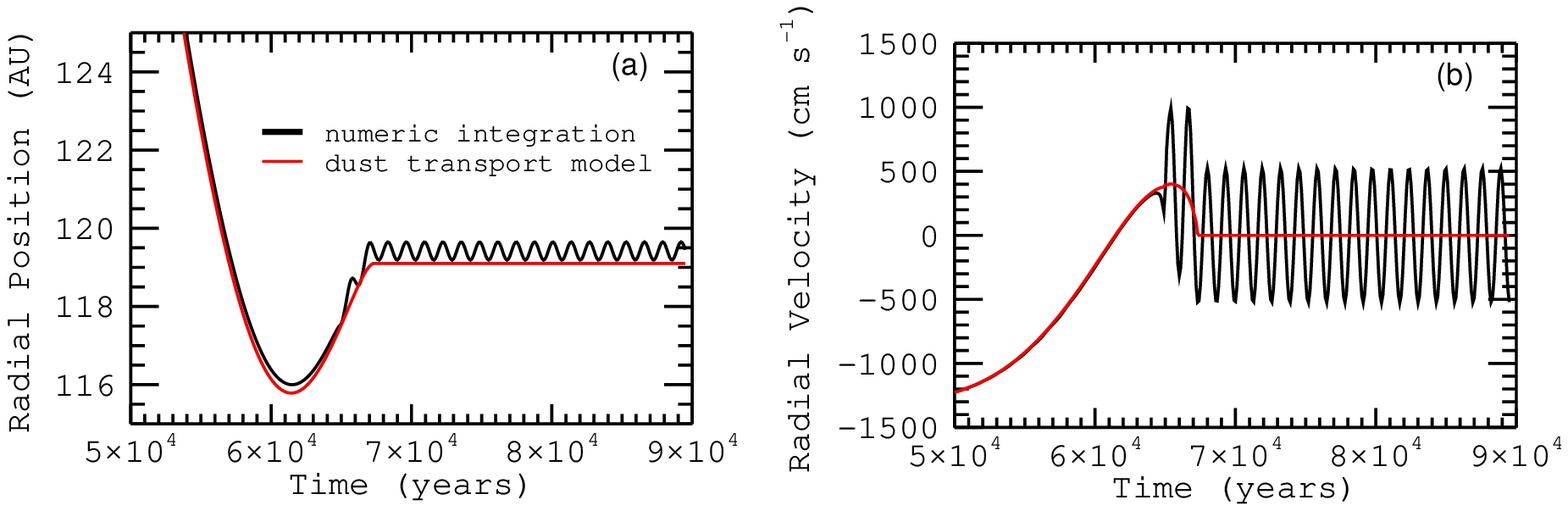}
   \caption{Radial position (a) and radial velocity (b) of a 2 mm particle initiated at 242 AU within a tenuous, evolving disk in the midplane-flow gas-velocity case.  Comparison between a precision numerical integration (black) and the particle-transport model trajectory (red).}
   \label{mvsAA2007horbit}
   \end{figure}

However, our particle-transport simulations do lack certain details; as shown in Figure \ref{mvsAA2007horbit}, 
they do not account for orbital eccentricity.  Thus, the particle-transport simulations are limited in their depiction of complex gas-particle interactions.  In Figure \ref{mvsAA2007horbit} these interactions pertain to a particle disengaging from and outward-sweeping photoevaporation front.  However, the results of the particle-transport simulations presented in this paper are not sensitive to such a high level of detail.  Instead, we are most interested in the broad effects of disk mass and particle size.  For the tenuous disk, we ran simulations for particles of sizes 0.2 $\mu$m, 2 $\mu$m, 20 $\mu$m, and 0.2 mm from 20 starting locations, 0.5--500 AU.  The exponential-stopping times in the simulations ranged from less than $10^{-4}$ to greater than $10$ Kepler times, and in general, the relative radial positions of the simulated trajectories agreed with the numeric integrations to within 2\% or better.

\begin{figure}
   \includegraphics[width=\textwidth]{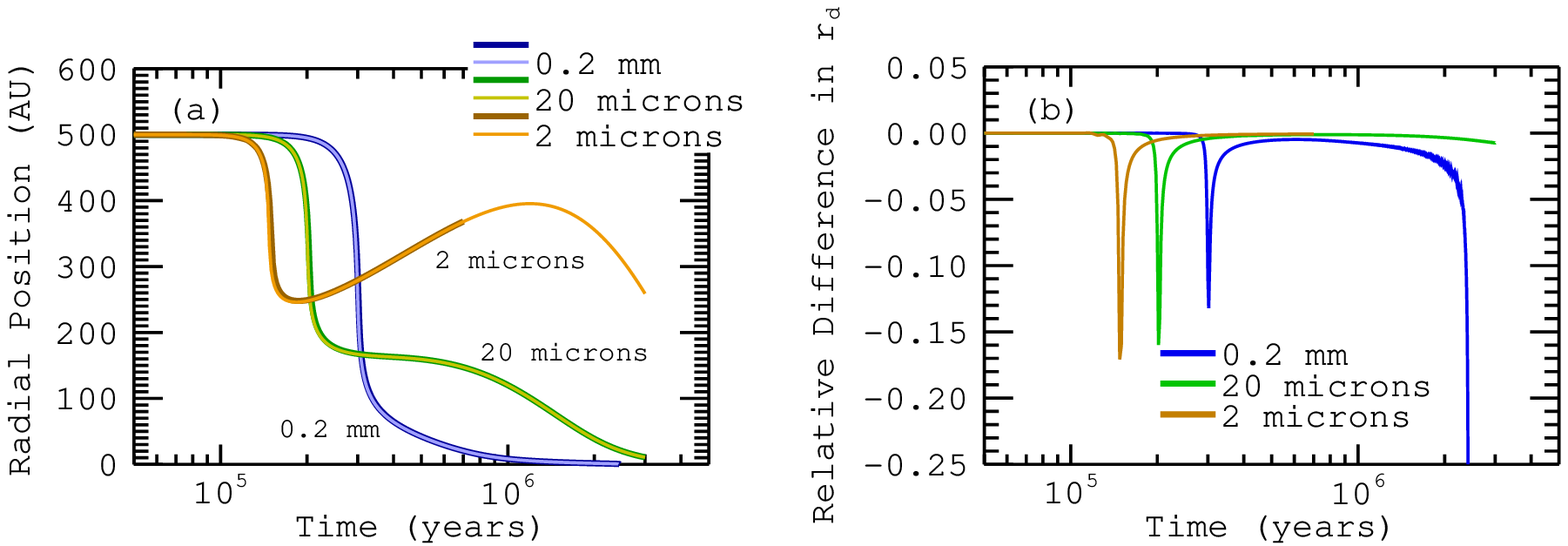}
   \caption{(a) Particle trajectories initiated at 500 AU within a massive, but initially compact disk for three particle sizes in the midplane-flow gas-velocity case.  Comparison between precision numerical integrations (heavy, dark lines) and particle-transport model trajectories (light lines). (b) Relative difference over time in the radial positions between the numerical integrations and the particle-transport simulations. [(simulated - integrated) / integrated]  Note that the time-axes are in log-scale.}
   \label{mvsM5R10longfall}
   \end{figure}

Where our particle-transport simulations deviate the most from the numeric integrations is for particles experiencing a long period of steady infall.  This is true regardless of how well coupled the particles are to the gas motions, and is an effect of the particle-transport model repeatedly overshooting the estimate of the infalling particle trajectory.  Figure~\ref{mvsM5R10longfall}(a) 
plots trajectory comparisons for particles initiated at 500 AU in a massive, but initially compact disk ($M_\mathrm{D}=0.05 M_\odot$ and $R_d=10$ AU).  In this scenario, the particles begin very decoupled from the gas ($T_s>10^3$ Kepler times), and after their initial infall toward the main disk they become moderately well coupled ($T_s\sim10^{-2}$ and smaller).  In Figure \ref{mvsM5R10longfall}(b) we plot the relative difference in the particle radial positions between the particle-transport model trajectories and the numerical integrations.  Here we can see that for each infall, these two trajectory calculations diverge as the simulated trajectories outpace those of the numeric integrations.  Therefore, exaggerated, rapid infall of particles may be a drawback of our particle-transport model.  However, inspection of all of our simulated trajectories suggests that this exaggerated infall should have a negligible effect on our general understanding of the timing and trends of particle transport.  As a check we have also run comparison integrations using the accretion-flow case in our nominal-disk model (Section \ref{Results_Flow}).  We ran four particle sizes (0.2 $\mu$m--0.2 mm) initiated at 15 AU and found positional differences of less than 2\% for all but the very final stages of infall. 

Therefore, in the absence of turbulence, the simulated trajectories produced by our particle-transport model appear to be a sufficiently accurate representation of radial particle transport in an evolving disk under the influence of gas drag, both from the pressure-supported--azimuthal and the radial gas flow.

     \subsection{Turbulent Diffusion Compared to Analytical Solutions}
     \label{Fidelity_Diffusion}

Here we demonstrate the fidelity of the random-walk method in our particle-transport model in reproducing accurate diffusion of the particle ensemble.  Using the analytical solutions of \cite{ClarkePringle1988} we compare the time-dependent particle distributions from our particle-transport simulations to the analytically expected contaminant distributions and discuss the effects of our choice of simulation time step on the results.  All solutions and simulations presented below use static-steady-disk--surface-density profiles.

In \cite{ClarkePringle1988}, the authors solve for the time-dependent concentration, $C$, of a contaminant initiated in a ring at specified radius: $C\left(t=0\right)=C_0\delta\left(R-R_0\right)$, where $C$ is the ratio of the local contaminant mass over the local disk mass, $R$ is the distance from the central star, and $R_0$ is the initial position of the contaminant.  The disk that they consider has a surface-density profile given by $\Sigma = \Sigma_0 R^{-a}$, where $\Sigma_0$ and $a$ are constants; it is also a steady disk, so that for disk viscosity, $\nu=\nu_0 R^b$ ($b$ and $\nu_0$ constants), $b=a$, and the gas-accretion velocity may be written as $v_R = v_0 R^{b-1}$ where $v_0 = -3\nu_0 / 2$.

Assuming azimuthal symmetry and that the gas and contaminant are vertically well mixed, the diffusion equation in polar coordinates may be written \citep{Gail2001}
\begin{equation}
  \frac{\partial}{\partial t}\left(\Sigma C\right)
+
  \frac{1}{R}\frac{\partial}{\partial R}\left(R\Sigma C v_R\right)
-
  \frac{1}{R}\frac{\partial}{\partial R}
  \left[R D_{ig}\Sigma\left(\frac{\partial C}{\partial R}\right)\right]
=
  P_i\left(R,t\right) \,,
\label{diffusioneq}
\end{equation}
where $D_{ig}$ is the coefficient of diffusion of the contaminant within the gas, and $P_i$ corresponds to the rate of production of the contaminant per radial increment within the disk.  \cite{ClarkePringle1988} (and the setup of our simulations) assume zero production of the contaminant, $P_i\left(R,t\right)=0$, and that the diffusivity scales with the disk viscosity, $D_{ig} = \zeta\nu$, where $\zeta$ is a constant.  For these particle-transport simulations using a static disk profile, Equation (\ref{diffusioneq}) is further constrained by $\partial\Sigma / \partial t=0$ and so $v_R$ refers specifically to the set dust-mean-radial velocity.  Using the profiles for the disk surface density and viscosity above, the full diffusion equation applicable to these simulations is:
\begin{equation}
  \frac{1}{\nu_0}\frac{\partial C}{\partial t}
+
  \frac{v_0}{\nu_0}R^{b-1}\frac{\partial C}{\partial R}
+
  \left(b-a\right)\frac{v_0}{\nu_0}R^{b-2}C
=
  R^{a-1}\frac{\partial}{\partial R}
         \left[\zeta R^{1+b-a}\left(\frac{\partial C}{\partial R}\right)\right]
\,,
\label{diffusioneqab}
\end{equation}
which simplifies to 
\begin{equation}
  \frac{1}{\nu_0}\frac{\partial C}{\partial t}
+
  \frac{v_0}{\nu_0}R^{a-1}\frac{\partial C}{\partial R}
=
  R^{a-1}\frac{\partial}{\partial R}
         \left[\zeta R\left(\frac{\partial C}{\partial R}\right)\right]
\,
\label{diffusioneqab88}
\end{equation}
in the steady-disk ($a=b$) scenario of \cite{ClarkePringle1988}.  The scenario of \cite{ClarkePringle1988} assumes an evolving (accreting) gas disk, but, because it is steady (has an unchanging relative--surface-density profile) and the diffusion equations do not depend on the actual value of the gas surface density, Equation (\ref{diffusioneqab88}) fits both that scenario and our static steady-disk scenario.

The general solution given in \cite{ClarkePringle1988} to Equation (\ref{diffusioneqab88}) using the initial contaminant distribution given above is
\begin{eqnarray}
    C\left(R,t\right)
&=&
    \frac{|q|C_0 R_0^{1-a}}{2 \zeta\nu_0 t}
    \left(\frac{R}{R_0}\right)^{v_0/2\zeta\nu_0}
    I_\beta\left(\frac{q^2 R_0^{1/q}R^{1/q}}{2\zeta\nu_0 t}\right)
    \exp\left[-q^2\frac{\left(R_0^{2-a}+R^{2-a}\right)}
                       {4\zeta \nu_0 t}
        \right] \,,
\nonumber \\
    q
&=&
    \frac{2}{\left(2-a\right)} \,,
\nonumber \\
    \beta
&=&
    \frac{|v_0/\zeta\nu_0|}{|2-a|} 
\,.
\label{solved}
\end{eqnarray}
where $I_\beta\left(x\right)$ is the modified Bessel function of the first kind of order $\beta$.

However, Equation (\ref{solved}) is not valid for the case of $a=b=2$, for which \cite{ClarkePringle1988} provide the solution
\begin{equation}
  C\left(R,t\right)
=
  \frac{C_0}{R_0}\left(4\pi\zeta\nu_0 t\right)^{-1/2}
  \exp\left(-\frac{\left[\ln\left(R/R_0\right)- v_0 t\right]^2}
                  {4\zeta\nu_0 t}\right) \,.
\label{aeqbeq2solved}
\end{equation}
Note, however, that Equation (\ref{aeqbeq2solved}) does not match Equation (3.1.3) in \cite{ClarkePringle1988} because of a typo in that paper, which accidentally presents the solution for twice the background velocity, $v_R \rightarrow 2v_R$.  Also, when $a=b=2$, the natural space to solve the equation is logarithmic space.  This happens to be the space used for the width of our model-disk grid cells, and may be related to the only odd behavior that we find in our particle transport simulations (see below).

From our particle-transport simulations, we output the number of particles per radial grid space.  Therefore, to compare the solutions provided by Equations~(\ref{solved}) and (\ref{aeqbeq2solved}) to our simulations, we must first convert the analytic $C\left(R,t\right)$ into an expected fractional-mass distribution.  The concentration can be expressed as the ratio of the contaminant-to-gas surface densities, $C=\sigma/\Sigma$, where $\sigma$ is the surface density of the contaminant.  Because these test cases assume a static-disk--surface-density profile, it is then relatively simple to solve for the fractional mass of the contaminant $m_i$, in a given grid space, where $i$ denotes the grid space, which we will normalize by $m_\mathrm{tot}$, the total mass of the contaminant in the disk at $t=0$, where
\[
  m_\mathrm{tot}
=
  2\pi \int_0^\infty R \Sigma C_0 \,\delta\left(R-R_0\right) dR
=
  2\pi R_0 \Sigma\left(R_0\right) C_0
\,.
\]
We write:
\begin{equation}
  \left(\frac{m_i}{m_\mathrm{tot}}\right)
=
  \left[ R_0 \Sigma\left(R_0\right) C_0 \right]^{-1}
  \int_{R=R_{i-1/2}}^{R_{i+1/2}}
        R \Sigma\left(R\right) C\left(R,t\right) dR \,,
\label{massfrac}
\end{equation}
where $R_{i-1/2}$ and $R_{i+1/2}$ are the inner and outer boundaries of the grid space, respectively.  Combining Equations~(\ref{solved}) and (\ref{massfrac}), and using a variable substitution, $s=R^{1/q}$, we can write the general expression for the expected mass distribution as
\begin{eqnarray}
    \left(\frac{m_i}{m_\mathrm{tot}}\right)
&=&
    \frac{q|q|R_0^{-v_0/2\zeta\nu_0}}{2\zeta\nu_0 t}
    \exp\left[-\frac{q^2R_0^{2-a}}{4\zeta\nu_0 t}\right]
    \int_{R=R_{i-1/2}}^{R_{i+1/2}}
        s^\gamma I_\beta\left(\frac{q^2s_0 s}{2\zeta\nu_0 t}\right)
        \exp\left(-\frac{q^2s^2}{2\zeta\nu_0 t}\right) ds
\nonumber \\
    \gamma
&=&
    \frac{\left(2-a+v_0/\zeta\nu_0\right)}{\left(2-a\right)}
\nonumber \\
   s_0
&=&
   R_0^{1/q} \,.
\label{mass_solved}
\end{eqnarray}
We solve the integral numerically using a Simpson integrator and the Numerical Recipes functions for the modified Bessel functions \citep{NumericalRecipes}.

For the case of $a=b=2$ given in Equation (\ref{aeqbeq2solved}) the expected mass distribution in a static disk is given by
\begin{equation}
\label{aeqbeq2solved_mass}
  \left(\frac{m_i}{m_\mathrm{tot}}\right)
=
  \left(4\pi\zeta\nu_0 t\right)^{-1/2}
  \int_{R=R_{i-1/2}}^{R_{i+1/2}}
      \exp\left[-\frac{\left(Y-v_0 t\right)^2}{4\zeta\nu_0 t}
          \right]
      dY \,,
\end{equation}
where $Y=\ln\left(R/R_0\right)$, and we again, we use a Simpson integrator to solve for the expected $\left(m_i/m_\mathrm{tot}\right)$ for these cases.

To compare our particle-transport model to the analytic test cases described in Equations~(\ref{mass_solved}) and (\ref{aeqbeq2solved_mass}), we ran simulations of 10,000 particles initiated at 40 AU; we include some simulations with zero background velocity, for which we substitute $v_0=0$ into the solution equations above (instead of the accretion-case $v_0=-3\nu_0/2$).  We vary the value of $a$, but all simulations use $\nu\left(1\mathrm{AU}\right) = 4.941\times10^{14}$~cm$^2$~s$^{-1}$ and $\zeta=1$.  In these test-case simulations, the values of $\Sigma_0$ and the particle size are arbitrary, since the background velocity of the particles, $v_R$, is prescribed.

\begin{figure}
   \includegraphics[width=\textwidth]{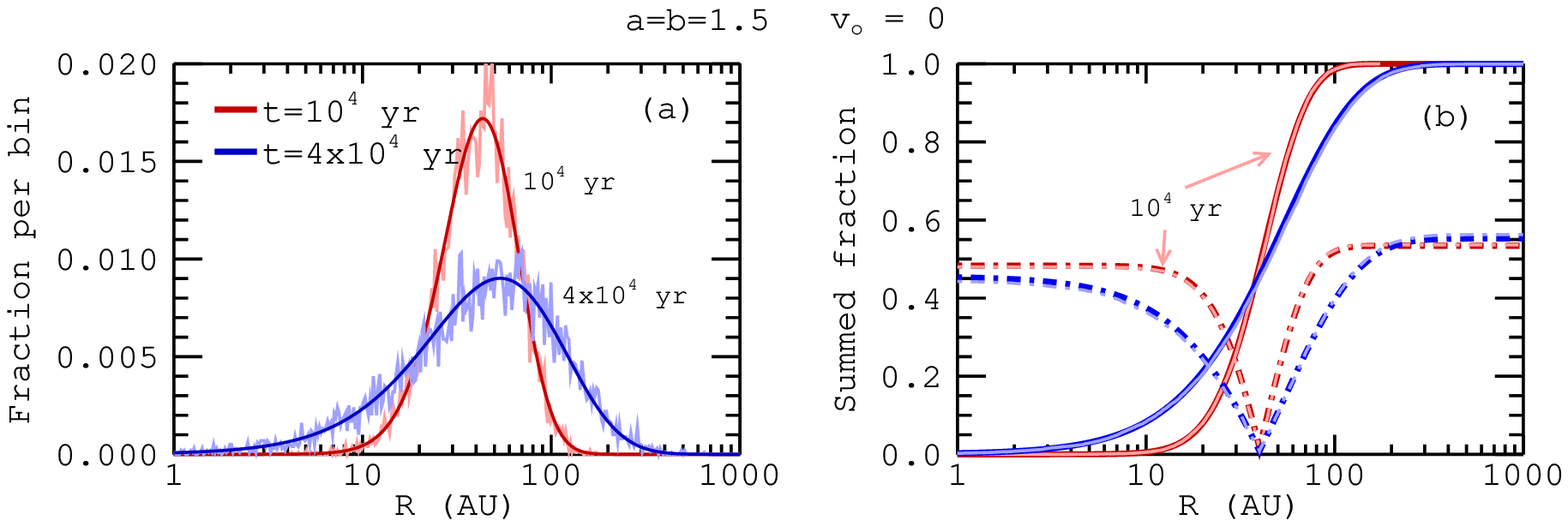}
   \caption{(a) Comparison of the mass distributions of the particles (contaminant) between the particle-transport simulations (pale lines) and the analytical solution (dark lines) at two times.  The vertical axis is the fraction of total particles (total contaminant mass) per radial grid space.  (b) Comparison between simulations (pale lines) and analytical solutions (dark lines) of the summed-mass distributions.  Solid lines represent the distributions summed from $R=0$, and dashed lines represent the distributions summed from $R=R_0=40$~AU.  For this figure $a=b=1.5$ and $v_0=0$.  For the simulations, 10,000 particles were initiated at 40 AU and diffused using a global time-step size of $\Delta t = 2000$ years.}
   \label{se15150}
   \end{figure}

Panel (a) of Figure~\ref{se15150} 
plots a comparison between our numerical simulations of particle transport and the analytically expected contaminant-mass distribution for the case of $a=1.5$ and zero background velocity at two different times. It shows that the overall, evolving distribution of the particle ensemble is in good agreement with the analytical solution.  However, the scatter produced by the discreet nature of the particle simulations makes it difficult to judge the accuracy with which the simulations reproduce this analytical test case.  Therefore, Figure \ref{se15150} panel (b) plots an alternate comparison of the simulated and analytical mass distributions.  Here the values plotted correspond to a summing of the fractional-mass per grid space either from $R=0$ outward, or from $R=R_0=40$~AU inward (for $R<R_0$) and outward (for $R>R_0$).  Again, the simulations and the analytical solutions are in good agreement.

\begin{figure}
   \includegraphics[width=\textwidth]{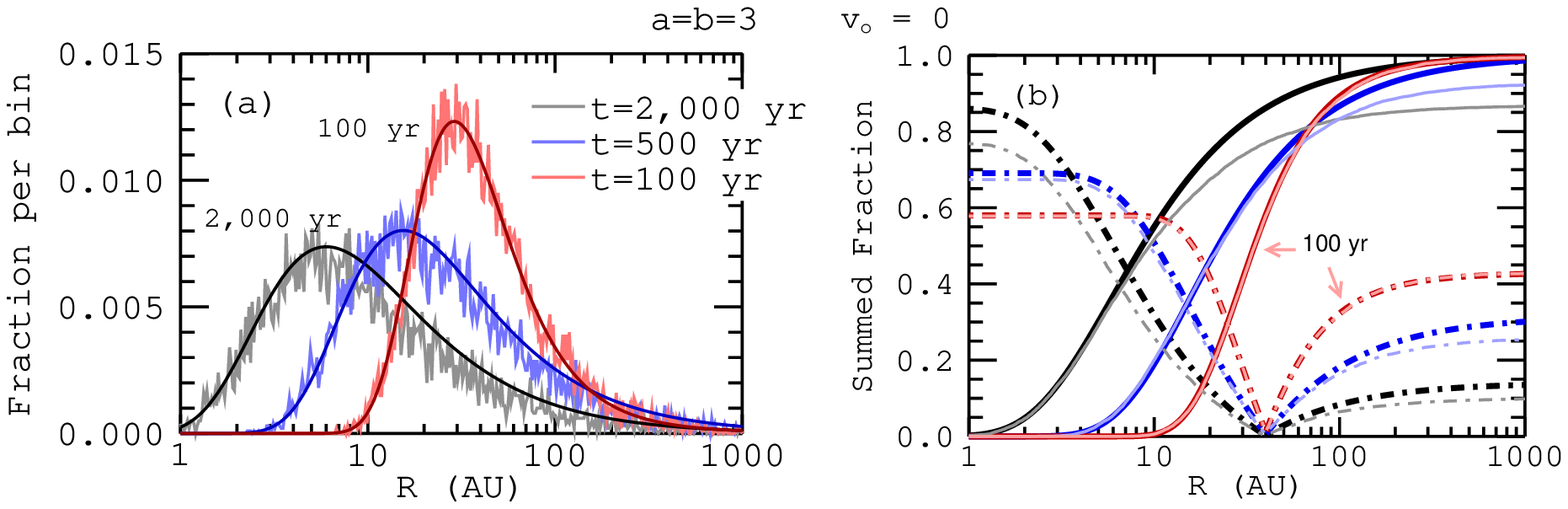}
   \caption{Mass-distribution profiles (a) and summed profiles (b) at three times for the case of $a=b=3$ with $v_0=0$.  Comparison between simulations (pale lines) and analytical solutions (dark lines).  In (b) solid lines represent summing from $R=0$, dashed lines from $R=R_0$.  In this simulation at $t=100$ years, 100\% of the particles remain on the grid, but at $t=500$ and 2000 years only 93\% and 87\% remain, respectively.  The simulation was run using a global time-step size of $\Delta t=100$ years.}
   \label{se330}
   \end{figure}

Simulating particle diffusion with zero background velocity in a steady disk for a range of $a$-values, we find, in general, that our simulations agree well with the analytical solutions.  The exceptions occur for very steep ($a=3$), or very shallow ($a=1$) disk profiles and are caused by the finite nature of our simulation space. A viscosity (diffusivity) profile that goes as $R^3$ leads to rapid transport of particles over very large distances at large $R$.  Therefore, the outer edge of our simulation space acts as a sink for particles.  At early times, before a significant fraction of particles have been lost, the simulations and analytical solutions are in good agreement, as shown in Figure \ref{se330}; 
they deviate from each other at later times.  In the $a=1$ case, it is loss past the inner boundary of the simulation space that causes the simulations to deviate from the analytical solutions.

\begin{figure}
   \includegraphics[width=\textwidth]{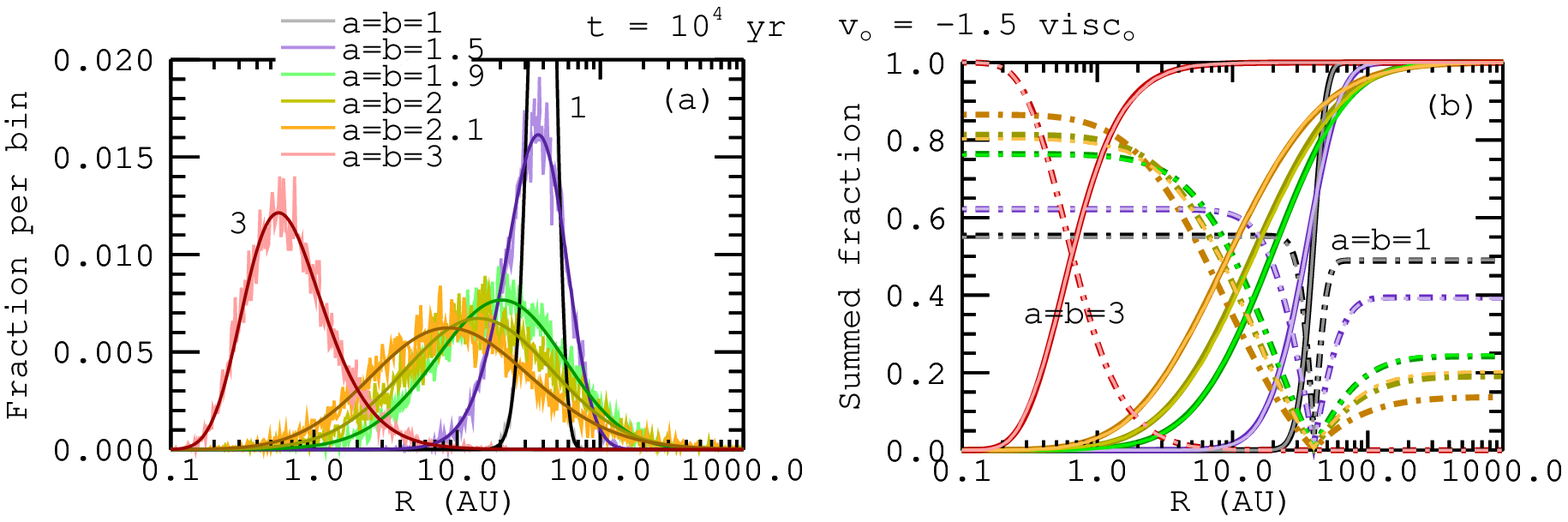}
   \caption{Mass-distribution profiles (a) and summed profiles (b) at $t=10^4$ years for several $a=b$ cases with $v_0=-1.5\nu_0$.  Comparison between simulations (pale lines) and analytical solutions (dark lines).  In (b) solid lines represent summing from $R=0$, dashed lines from $R=R_0$. For the case of $a=b=1$, the simulation was run using a global time-step size of $\Delta t=10^4$ years.  For $a=b=$ 1.5 and 3, $\Delta t=1000$ years.  For $a=b=$ 1.9, 2, and 2.1, $\Delta t=500$ years.}
   \label{aeqbVacc}
   \end{figure}

Loss of simulation particles past the simulation-space boundaries is not an issue when we include the background-accretion velocity.  In Figure \ref{aeqbVacc}, 
we plot the mass-distribution profiles at $t=10^4$ years for a range of $a$-values with the accretion velocity included.  Because $v_R$ is inward, it keeps particles away from the outer boundary, and loss past the inner boundary is a part of the analytical solution as well as the simulations.  All of these cases show good agreement between the simulations and the analytical solutions.  

\begin{figure}
   \includegraphics[width=\textwidth]{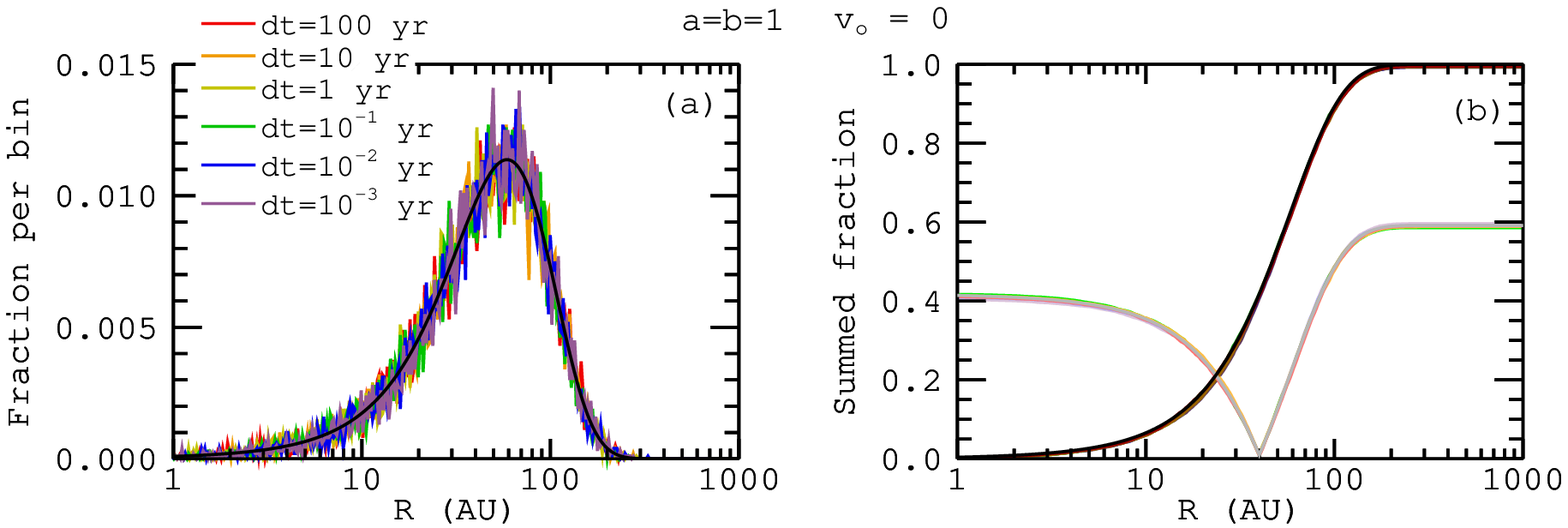}
   \caption{Mass-distribution profiles (a) and summed-profiles (b) at $t=2\times10^5$ years for the steady-disk case where $a=b=1$ with $v_0=0$.  Simulations were run for a range of global time-step sizes (colored lines) and are compared to the analytical solution (black/grey lines).  In (b) the dark lines represent summing from $R=0$, and the pale lines summing from $R=R_0$.  The colored lines of the simulations are largely obscured by the analytical solutions (black/grey) plotted over top.}
   \label{dt110}
   \end{figure}

Finally, we would like to be sure that our particle-transport model correctly simulates diffusion of the particle ensemble for the global time steps used in our static-disk models ($\Delta t=2000$ years), and for the time steps required by the disk-evolution model ($\Delta t\sim0.05$ years).  As was described in Section \ref{Transport}, maximum limits are placed on the particle time stepping to ensure that disk conditions local to each particle are obeyed.  However, smaller time steps mean more steps to simulate a given amount of time in the disk; therefore, the probabilities of a particle stepping inward versus outward are closer and closer to equal.  Cumulatively, however, all time-step sizes used must evolve the distribution of the particle ensemble equally.  Therefore, we have run a few test-case simulations using time-step sizes ranging from $10^{-3}$ years to 100 years.  For the case of $a=1$ and $v_0=0$, the resultant distributions versus the analytical solutions are plotted in Figure \ref{dt110}. 
All of these simulations agree well with the analytical solution, and the scatter produced for simulations with different time-step sizes are all comparable.

\begin{figure}
   \includegraphics[width=\textwidth]{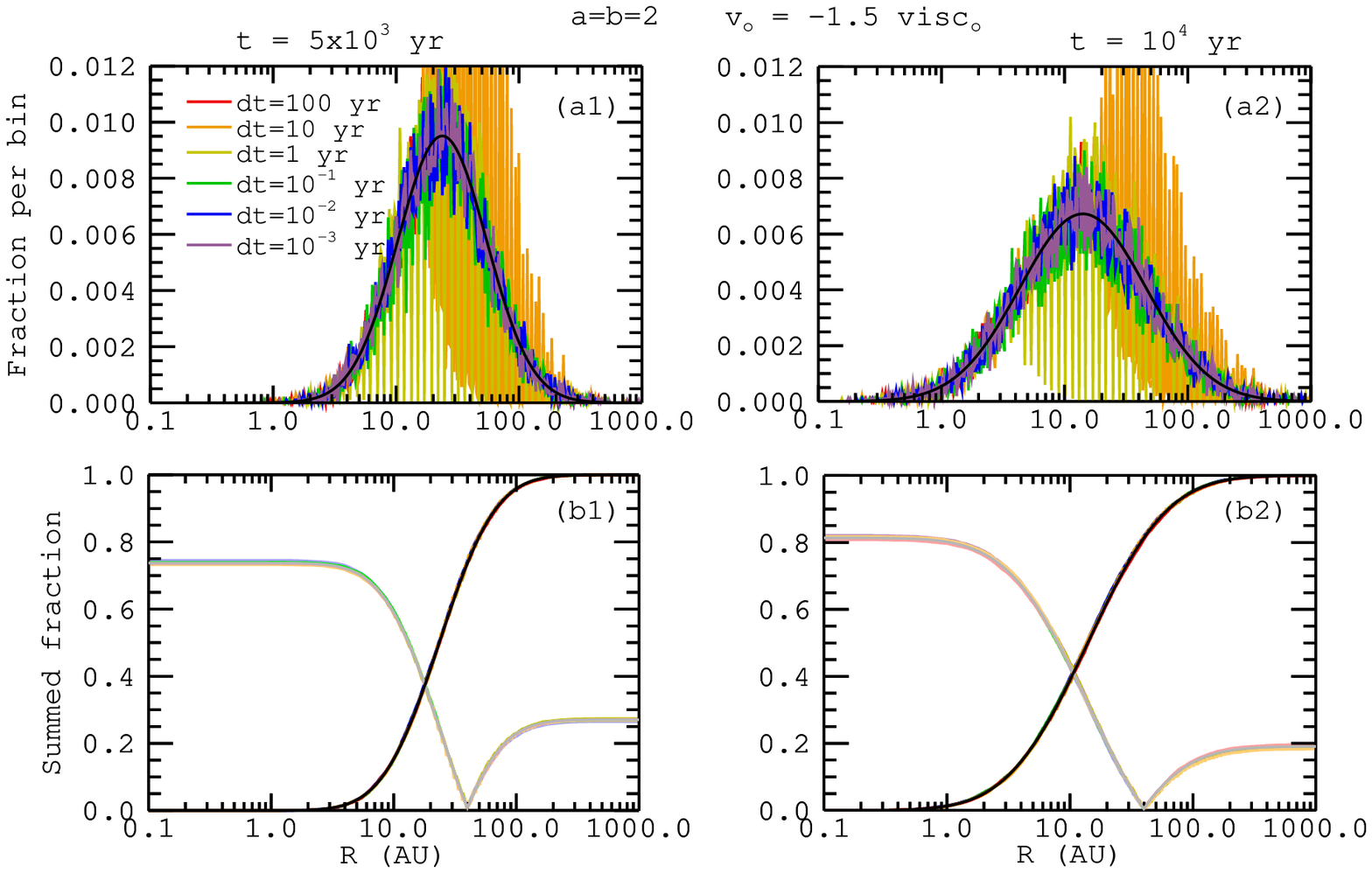}
   \caption{Mass-distribution profiles (a1 and a2) and summed-profiles (b1 and b2) at $t=5\times10^3$ years (a1 and b1) and $t=10^4$ years (a2 and b2) for the steady-disk case where $a=b=2$, and $v_0=-1.5\nu_0$.  Simulations were run for a range of global time-step sizes (colored lines) and are compared to the analytical solution (black/grey lines).  In (b1) and (b2), the dark lines represent summing from $R=0$, and the pale lines summing from $R=R_0$.  The colored lines of the simulations are largely obscured by the analytical solutions (black/grey) plotted over top.}
   \label{dt221}
   \end{figure}

While the agreement between simulations using different time-step sizes is good for almost all of the simulations we tested, one set of simulations shows a marked change in the scatter depending on the size of the time step used.  Figure~\ref{dt221} 
plots simulations run for a range of time-step sizes for the case of $a=b=2$ and $v_0=-3\nu_0/2$ compared to the analytical solutions at two times.  Looking at the summed-mass distributions, it is clear that all of these simulations do distribute the particle ensemble appropriately for this diffusion test case.  However, the mass-fraction-per-bin profiles show significant, resonant-like scatter in the $\Delta t=10$ years and $\Delta t=1$ year simulations; the scatter also appears a bit higher than usual in the $\Delta t=10^{-1}$ year simulation.  While it is not clear what causes this large scatter for these time steps, it is suggestive that the natural space for diffusion in this regime ($a=b=2 \longrightarrow \ln\left(R\right)$-space) is the same as the space wherein our disk-model grid cells are equal width.  Furthermore, the resonant-like-scatter region in the $\Delta t=10$ years simulation is restricted to outward of about 15~AU, and this region moves outward with time.  Inward of 15~AU in these simulations, the Kepler time ($1/\Omega_\mathrm{K}$) is less than 10 years and decreases toward smaller $R$; the time-step sizes of individual particles at small AU are restricted to smaller than the global 10-year time step.  Therefore, the resonant-like-scatter behavior is only observable when the same local time-step size is used for the entire region and when the source of all the particles is a single point in $R$.  Nonresonant-like scatter at small AU allows the particle distribution to spread out randomly, following the local diffusivity and removing the single-source-point condition starting at small AU and moving outward.

While this resonant-like-scatter behavior is curious and somewhat vexing, it is not a problem for the simulations presented in this paper.  Aside from the fact that our model disks (Section \ref{Disk_Evolution}) use $\nu=\nu_0 R$ ($b=1$), the time-step sizes for our steady-disk and static0-disk models are large enough that the local time-step restrictions of the particle-transport model apply almost everywhere.  The evolving-disk models use a time-step size that is small enough ($\Delta t\sim 0.05$ years) that resonant-like scatter is not expected to be an issue. Note, though, that resonant-like-scatter behavior may be something to watch out for when using this method of particle-transport modeling.


\begin{thebibliography}{999}

\bibitem[Alexander \& Armitage(2007)]{AlexanderArmitage2007}
Alexander, R.~D., \& Armitage, P.~J., 2007, \mnras, 375, 500

\bibitem[Bell et al.(1997)]{Belletal1997}
Bell, K.~R., Cassen, P.~M., Klahr, H.~H., \& Henning, Th. 1997, \apj, 486, 372

\bibitem[Bockel\'ee-Morvan et al.(2002)]{BockeleeMorvanetal2002}
Bockel\'ee-Morvan, D., Gautier, D., Hersant, F., Hur\'e, J.-M., \& Robert, F. 2002,  A\&A, 384, 1107 

\bibitem[Brownlee et al.(2006)]{Brownleeetal2006}
Brownlee, D., et al. 2006, Science, 314, 1711

\bibitem[Boss(2004)]{Boss2004}
Boss, A.~P. 2004, \apj, 616, 1265

\bibitem[Boss(2008)]{Boss2008} 
Boss, A.~P. 2008, Earth \& Planetary Science, 268, 102

\bibitem[Ciesla(2007)]{Ciesla2007}
Ciesla, F.~J. 2007, Science, 318, 613

\bibitem[Ciesla(2009)]{Ciesla2009}
Ciesla, F.~J. 2009, Icarus, 200, 655

\bibitem[Clarke \& Pringle(1988)]{ClarkePringle1988}
Clarke, C.~J., \& Pringle, J.~E. 1988, \mnras, 235, 365

\bibitem[Clarke \& Lodato(2009)]{ClarkeLodato2009}
Clarke, C.~J., \& Lodato, G. 2009, \mnras, 398, L6

\bibitem[Desch(2007)]{Desch2007}
Desch, S.~J. 2007, \apj, 671, 878

\bibitem[Dullemond, Apai, \& Walch(2006)]{DullemondApaiWalch2006}
Dullemond, C.~P., Apai, D., \& Walch, W. 2006, \apjl, 640, L67

\bibitem[Dullemond, Natta, \& Testi(2006)]{DullemondNattaTesti2006}
Dullemond, D.~P., Natta, A., \& Testi, L. 2006, \apjl, 645, L69

\bibitem[Dullemond et al.(2008)]{Dullemondetal2008}
Dullemond, C.~P., Brauer, F., Henning, Th, \& Natta, A. 2008, Physica Scripta, 130, 014015

\bibitem[Font et al.(2004)]{Fontetal2004}
Font, A. S., McCarthy, I. G., Johnstone, D., \& Ballantyne, D. R. 2004, \apj, 607, 890

\bibitem[Gail(2001)]{Gail2001}
Gail, H.-P. 2001, A\&A, 378, 192

\bibitem[Gail(2004)]{Gail2004}
Gail., H.-P. 2004, A\&A, 413, 571

\bibitem[Garaud(2007)]{Garaud2007}
Garaud, P. 2007, \apj, 671, 2091

\bibitem[Hanner et al.(1994)]{Hanneretal1994}
Hanner, M.~S., Hackwell, J.~A., Russell, R.~W., \& Lynch, D.~K. 1994, Icarus, 112, 490

\bibitem[Harker \& Desch(2002)]{HarkerDesch2002}
Harker, D.~E., \& Desch, S.~J. 2002, \apj, 565, L109

\bibitem[Harker et al.(2002)]{Harkeretal2002}
Harker, D.~E., Wooden, D.~H., Woodward, C.~E., \& Lisse, C.~M. 2002, \apj, 580, 579

\bibitem[Hartmann et al.(1998)]{Hartmannetal1998}
Hartmann, L., Calvet, N., Gullbring, E., \& D'Alessio, P. (1998) \apj, 495, 385

\bibitem[Hollenbach et al.(1994)]{Hollenbachetal1994}
Hollenbach, D., Johnstone, D., Lizano, S. \& Shu, F. 1994, \apj, 428, 654

\bibitem[Honda et al.(2004)]{Hondaetal2004}
Honda, M., et al. 2004, \apj, 601, 577

\bibitem[Hueso \& Guillot(2005)]{HuesoGuillot2005}
Hueso, R., \& Guillot, T. 2005, A\&A, 442, 703

\bibitem[Joswiak et al.(2010)]{Joswiaketal2010}
Joswiak, D.~J., Brownlee, D.~E., Matrajt, G., Messenger, S.~M., \& Ito, M. 2010, 41st Lunar and Planetary Sciences Conference, no.~2119

\bibitem[Keller \& Gail(2004)]{KellerGail2004} 
Keller, Ch., \& Gail, H.-P. 2004, A\&A, 414, 1177

\bibitem[Kenyon \& Hartmann(1987)]{KenyonHartmann1987}
Kenyon, S.~J., \& Hartmann, L. 1987, \apj, 323, 714

\bibitem[King, Pringle, \& Livio(2007)]{KingPringleLivio2007}
King, A.~R., Pringle, J.~E., \& Livio, M. 2007, \mnras, 376, 1740

\bibitem[Krauss \& Wurm(2005)]{KraussWurm2005}
Krauss, O., \& Wurm, G. 2005, \apj, 630, 1088

\bibitem[Lin \& Pringle(1990)]{LinPringle1990}
Lin, D.~N.~C., \& Pringle, J.~E. 1990, \apj, 358, 515

\bibitem[Lisse et al.(2006)]{Lisseetal2006}
Lisse, C.~M., et al. 2006, Science, 313, 635

\bibitem[Lodders(2003)]{Lodders2003}
Lodders, K. 2003, \apj, 591, 1220

\bibitem[Molster \& Kemper(2005)]{MolsterKemper2005}
Molster, F., \& Kemper, C. 2005, Space Sci.~Rev., 119, 3

\bibitem[Morfill \& V\"olk(1984)]{MorfillVolk1984}
Morfill, G. E., \& V\"olk, H.~J. 1984, \apj, 287, 371

\bibitem[Mousis et al.(2007)]{Mousisetal2007}
Mousis, O., Petit, J.-M., Wurm, G., Krauss, O., Alibert, Y., \& Horner, J. 2007, A\&A, 466, L9.

\bibitem[Nuth \& Johnson(2006)]{NuthJohnson2006}
Nuth, J.~A. III, \& Johnson, N.~M. 2006, Icarus, 180, 243

\bibitem[Olofsson et al.(2009)]{Olofssonetal2009}
Olofsson, J., et al. 2009, A\&A, 507, 327

\bibitem[Owen et al.(2009)]{Owenetal2009}
Owen, J.~E., Ercolano, B., Clarke, C.~J., \& Alexander, R.~D. 2009, \mnras, doi:10.1111/j.1365-2966.2009.15771.x

\bibitem[Pavlyuchenkov \& Dullemond(2007)]{PavlyuchenkovDullemond2007}
Pavlyuchenkov, Ya., \& Dullemond, C.~P. 2007, A\&A, 471, 833-840

\bibitem[Press et al.(1992)]{NumericalRecipes}
Press, W.~H., Teukolsky, S.~A., Vetterling, W.~T., \& Flannery, B.~P. 1992, Numerical Recipes in Fortran 77: Second Edition: The Art of Scientific Computing, available online: http://www.nrbook.com/a/bookfpdf.php, \S 6.6 and \S 6.7

\bibitem[Pringle(1981)]{Pringle1981}
Pringle, J.~E. 1981, ARA\&A, 19, 137

\bibitem[Pringle, Verbunt \& Wade(1986)]{PringleVerbuntWade1986} 
Pringle, J.~E., Verbunt, F., \& Wade, R.~A. 1986, \mnras, 221, 169

\bibitem[Rozyczka, Bodenheimer \& Bell(1994)]{RozyczkaBodenheimerBell1994}
Rozyczka, M., Bodenheimer, P., \& Bell, K. R. 1994, \apj, 423, 736

\bibitem[Scott \& Krot(2005)]{ScottKrot2005}
Scott, E.~R.~D., \& Krot, A.~N. 2005, \apj, 623, 571

\bibitem[Shakura \& Sunyaev(1973)]{ShakuraSunyaev1973} 
Shakura, N.~I., \& Sunyaev, R.~A. 1973, A\&A, 24, 337

\bibitem[Shu et al.(2001)]{Shuetal2001}
Shu, F.~H., Shang, H., Gounelle, M., Glassgold, A.~E., \& Lee, T. 2001, \apj, 548, 1029

\bibitem[Stodolna, Jacob \& Leroux(2010)]{StodolnaJacobLeroux2010}
Stodolna, J., Jacob, D., \& Leroux, H. 2010, 41st Lunar and Planetary Sciences Conference, no.~1657

\bibitem[Takeuchi \& Lin(2002)]{TakeuchiLin2002}
Takeuchi, T., \& Lin, D.~N.~C. 2002, \apj, 581, 1344

\bibitem[Tscharnuter et al.(2009)]{Tscharnuteretal2009}
Tscharnuter, W.~M., Sch\"onke, J., Gail, H.-P., \& L\"uttjohann, E. 2009, A\&A, 504, 109

\bibitem[Turner, Carballido, \& Sano(2010)]{TurnerCarballidoSano2010}
Turner, N.~J., Carballido, A., \& Sano, T. 2010, \apj, 708, 188

\bibitem[Urpin(1984)]{Urpin1984}
Urpin, V.~A. 1984, Soviet Astronomy, 28, 50

\bibitem[van Boekel et al.(2004)]{vanBoekeletal2004}
van Boekel, R., et al. 2004, Nature, 432, 479

\bibitem[van Boekel et al.(2005)]{vanBoekeletal2005}
van Boekel, R., Min, M., Waters, L.~B.~F.~M., de Koter, A., Dominik, C., van den Ancker, M.~E., \& Bouwman, J. 2005, A\&A, 437, 189

\bibitem[Vinkovi\'c(2009)]{Vinkovic2009}
Vinkovi\'c, D. 2009, Nature, 459, 227

\bibitem[Watson et al.(2009)]{Watsonetal2009}
Watson, D.M., et al. 2009, \apjs, 180, 84

\bibitem[Weidenschilling(1977a)]{Weidenschilling1977a}
Weidenschilling, S.~J. 1977a, \mnras, 180, 57

\bibitem[Weidenschilling(1977b)]{Weidenschilling1977b}
Weidenschilling, S.~J. 1977b, Ap\&SS, 51, 153

\bibitem[Westphal et al.(2009)]{Westphaletal2009}
Westphal, A.~J., Fakra, S.~C., Gainsforth, Z., Marcus, M.~A., Ogliore, R.~C., \& Butterworth, A.~L. 2009, \apj, 694, 18

\bibitem[Wooden, Woodward, \& Harker(2004)]{WoodenWoodwardHarker2004}
Wooden, D.~H., Woodward, C.~E., \& Harker, D.~E. 2004, \apjl, 612, L77

\bibitem[Wooden, Harker, \& Brearley(2005)]{WoodenHarkerBrearley2005}
Wooden, D.~H., Harker, D.~E., \& Brearley, A.~J. 2005, ASP Conference Series, 341, 774

\bibitem[Wooden et al.(2007)]{Woodenetal2007}
Wooden, D., Desch, S., Harker, D., Gail, H.-P., \& Keller, L. 2007, in Protostars and Planets V, eds. B. Reipurth, D. Jewitt, \& K. Keil (University of Arizona Press: Tuscon) 815

\bibitem[Wooden(2008)]{Wooden2008}
Wooden, D.~H. 2008, Space Sci.~Rev., 138, 75

\bibitem[Youdin \& Lithwick(2007)]{YoudinLithwick2007} 
Youdin, A.~N., \& Lithwick, Y. 2007, Icarus 192, 588

\bibitem[Zolensky et al.(2006)]{Zolenskyetal2006}
Zolensky, M.~E., et al. 2006, Science, 314, 1735

\end{thebibliography}
\end{document}